\documentclass[a4paper,journal]{IEEEtran}  

\usepackage{multirow}

\usepackage{color}
\usepackage{float}
\usepackage{amsmath}
\usepackage{amssymb}
\usepackage{mathrsfs}
\usepackage{mathtools}

\usepackage{amsthm}
\newtheorem{theorem}{Theorem}[]
\newtheorem{lemma}[theorem]{Lemma}

\usepackage{graphicx}
\makeatletter

\providecommand{\tabularnewline}{\\}

\usepackage[ruled,vlined,linesnumbered]{algorithm2e}
\SetKwInput{KwInput}{Input}
\SetKwInput{KwOutput}{Output}

\usepackage{algorithmicx}
\usepackage{algpseudocode}
\usepackage{amsmath,amsfonts,amssymb}
\usepackage{amssymb}

\usepackage{array,booktabs,arydshln}
\usepackage{arydshln}

\usepackage{float} % Force figure position

\usepackage[caption=false,font=footnotesize]{subfig}

\@ifundefined{showcaptionsetup}{}{%
 \PassOptionsToPackage{caption=false}{subfig}}

\usepackage{subfig}
\usepackage{adjustbox}
\usepackage{multicol}
\usepackage{multirow}

\makeatother

\usepackage[dvipsnames,table]{xcolor} 

\definecolor{mdarkblue}{rgb}{0.0,0.0,0.6}
\definecolor{mdarkblue2}{RGB}{0,26,153}
\definecolor{mdarkred}{RGB}{174,49,54}
\definecolor{Prune}{RGB}{99,0,60}

\definecolor{orange0}{RGB}{252,169,133}
\definecolor{orange1}{RGB}{253,202,162}
\definecolor{orange2}{RGB}{254,235,201}
\definecolor{yellow0}{RGB}{255,237,81}
\definecolor{yellow1}{RGB}{255,250,129}
\definecolor{yellow2}{RGB}{255,255,176}

\definecolor{green0}{RGB}{133,202,93}
\definecolor{green1}{RGB}{191,228,118}
\definecolor{green2}{RGB}{224,243,176}

\definecolor{cyan0}{RGB}{111,183,214}
\definecolor{cyan1}{RGB}{154,206,223}
\definecolor{cyan2}{RGB}{204,236,239}

\definecolor{violet0}{RGB}{165,137,193}
\definecolor{violet1}{RGB}{193,179,215}
\definecolor{violet2}{RGB}{221,212,232}

\definecolor{pink0}{RGB}{249,140,182}
\definecolor{pink1}{RGB}{251,182,209}
\definecolor{pink2}{RGB}{253,222,238}

\newcommand{\cG}{\mathcal{G}}
\newcommand{\cN}{\mathcal{N}}
\newcommand{\cL}{\mathcal{L}}
\newcommand{\cK}{\mathcal{K}}
\newcommand{\cS}{\mathcal{S}}
\newcommand{\cP}{\mathcal{P}}

\newcommand{\abilene}{\textsc{Abilene}}

\newcommand{\costnet}{\textsc{Cost266}}
\newcommand{\twofat}{$2$-ary fat-tree}
\newcommand{\sixfat}{$6$-ary fat-tree}

\newcommand{\ilpse}{$\mathtt{ILP}$}
\newcommand{\bnbstar}{$\mathtt{BnB^*}$}
\newcommand{\bnbstarinf}{$\mathtt{BnB^*}\text{-}\infty$}
\newcommand{\bnbstarlim}[1]{$\mathtt{BnB^*}\text{-}#1$}
\newcommand{\bfnse}{$\mathtt{BFN}$}
\newcommand{\hidden}[1]{}

\usepackage{cite}

\begin{document}

\markboth{Preprint}%
{Luu \MakeLowercase{\textit{et al.}}: Network Slicing with Flexible VNF Order: A Branch-and-Bound Approach}

\title{Network Slicing with Flexible VNF Order: A Branch-and-Bound Approach}

\author{
    Quang-Trung~Luu,
    Minh-Thanh Nguyen,
    Tuan-Anh Do,
    Michel Kieffer,
    Van-Dinh Nguyen,\\
    Tai-Hung Nguyen, and
    Huu-Thanh Nguyen% <-this % stops a space
    \thanks{Quang-Trung~Luu,
    Minh-Thanh Nguyen,
    Tuan-Anh Do,
    Tai-Hung Nguyen, and
    Huu-Thanh Nguyen are with School of Electrical and Electronic Engineering, Hanoi University of Science and Technology, Hanoi 100000, Vietnam
    (email: trung.luuquang@hust.edu.vn). Corresponding author: Quang-Trung Luu (trung.luuquang@hust.edu.vn).}%
    \thanks{Michel Kieffer is with Universit\'e Paris-Saclay -- CNRS -- CentraleSup\'elec -- L2S, Gif-sur-Yvette, F-91192, France
    (email:  michel.kieffer@l2s.centrale supelec.fr).}% 
    \thanks{Van-Dinh Nguyen is with College of Engineering and Computer Science \& Center for Environmental Intelligence, VinUniversity, Vinhomes Ocean Park, Hanoi 100000, Vietnam
    (email: dinh.nv2@vinuni.edu.vn).}%  
    \thanks{This research is funded by Hanoi University of Science and Technology (HUST) under project number T2023-TD-011.}%
}%

\maketitle

\begin{abstract}
Network slicing is a critical feature in 5G and beyond communication systems, enabling the creation of multiple virtual networks (\emph{i.e.}, \emph{slices}) on a shared physical network infrastructure. This involves efficiently mapping each slice component, including virtual network functions (VNFs) and their interconnections (virtual links), onto the physical network. 

This paper considers slice embedding problem in which the order of VNFs can be adjusted, providing increased flexibility for service deployment on the infrastructure. This also complicates embedding, as the best order has to be selected. We propose an innovative optimization framework to tackle the challenges of jointly optimizing slice admission control and embedding with flexible VNF ordering. Additionally, we introduce a near-optimal branch-and-bound (BnB) algorithm, combined with the A* search algorithm, to generate embedding solutions efficiently. Extensive simulations on both small and large-scale scenarios demonstrate that flexible VNF ordering significantly increases the number of deployable slices within the network infrastructure, thereby improving resource utilization and meeting diverse demands across varied network topologies. 
\end{abstract}


\IEEEpeerreviewmaketitle{}

%%%%%%%%%%%%%%%%%%%%%%%%%%%%%%%%%%%%%
\section{Introduction} 
\label{sec:Introduction}
\IEEEPARstart{N}{etwork} slicing is a transformative technology within 5G and beyond communication systems that enables the dynamic creation of multiple, isolated virtual networks on a shared physical infrastructure \cite{foukas2017network,rafique2024survey,ebrahimi2024resource}. Each of these virtual networks, known as \emph{slices}, is carefully configured to meet the diverse and stringent requirements of various applications and services, which may range from high-bandwidth video streaming to ultra-reliable, low-latency applications such as industrial automation and autonomous vehicles. This customization is achieved through the allocation of dedicated network resources to ensure the necessary performance characteristics for each slice, in terms of bandwidth, latency, reliability, and security \cite{su2019resource,dealwis2023survey,donatti2023survey}. 

% With the advent of 5G and the forward-looking development of 6G, the role of network slicing has become even more crucial, as it empowers network operators to accommodate the rapidly diversifying requirements of contemporary digital services efficiently, all while operating on a single, unified physical network infrastructure. Network slicing thus not only enhances resource utilization but also opens avenues for innovation by allowing service providers to tailor network functionalities and performance, driving significant advancements in how networks handle the growing demands of modern communication ecosystems.

A key challenge in network slicing is the problem of network slice embedding (NSE) \cite{rafique2024survey,ebrahimi2024resource}. 
This involves mapping virtual network functions (VNFs) and their interconnections (called \emph{virtual links}) onto the physical network infrastructure in an optimized manner. VNFs are software-based network functions that replace traditional hardware components, providing increased flexibility and scalability. Successfully embedding these VNFs requires careful consideration of resource availability, network topology, and service level agreements (SLAs) \cite{saha2024survey,zangooei2024flexible}. 
An optimal embedding maximizes efficient resource utilization, meets the performance requirements of each slice, and minimizes operational costs. However, finding this optimal embedding is computationally complex, especially within large-scale, dynamic network environments \cite{foukas2017network,su2019resource}.

Most prior work on network slicing assumes a fixed slice structure, meaning that the VNFs of each slice are chained in a predetermined order \cite{Riggio2016, bouten2017semantically, Vizarreta2018qos, luu2022admission, Zhang2023multi, thanh2024accelerating}. 
This approach overlooks the potential advantages and challenges of flexible VNF ordering. In practice, multiple slice structure variants can be considered to deliver the same service. These variants might involve different sequences of VNFs or even alternative types of VNFs within the slice composition \cite{ocampo2017optimal, huang2015converged}.
\begin{figure}[t]
    \centering
    \includegraphics[width=0.8\columnwidth]{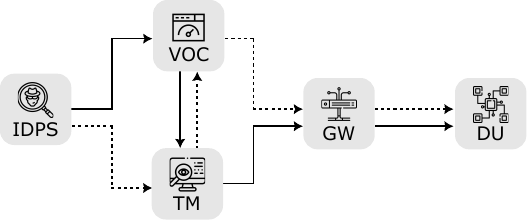}
    \caption{An example of a slice with flexible VNF ordering is one dedicated to a video streaming service, consisting of five VNFs: an intrusion detection and prevention system (IDPS), video optimization controller (VOC), traffic monitoring (TM), gateway (GW), and distributed unit (DU). In this case, the TM and VOC functions can be swapped with each other, resulting in two possible slice configurations. \label{fig:flexible-example}}
\end{figure}

Fig.~\ref{fig:flexible-example} illustrates a slice with flexible VNF ordering, adapted from the example slice in \cite{luu2020coverage}. This slice supports video streaming service at downlink and  
has five VNFs, including an intrusion detection and prevention system (IDPS), a video optimization controller (VOC), a traffic monitoring (TM), a gateway (GW), and a distributed unit (DU). The IDPS, GW, and DU functions are placed at fixed positions $1$, $2$, and $5$, respectively. The other VNFs, VOC and TM, can be flexibly placed at the third and fourth positions.
Additionally, different resource requirements and quality of service may be associated with each of these variants. 
The ability to select from multiple slice configurations to implement a service offers the virtual network operator (VNO) added flexibility during slice embedding.

\vspace{0.1cm}
\noindent{\bf Contributions.}
This paper addresses the challenges of optimizing network slicing in 5G networks and beyond by introducing a novel framework that allows flexible VNF ordering. 
The contributions of this paper are summarized as follows.
\begin{itemize}
    \item First, we propose a novel optimization framework that jointly tackles the challenges of ($i$) slice admission control, ($ii$) VNF order selection, and ($iii$) slice embedding. Unlike traditional methods that adhere to predetermined VNF sequences, this framework allows network operators to dynamically rearrange the order of VNFs within each slice, optimizing for performance and resource utilization. This results in a nonlinear integer programming problem;

    \item To facilitate the solution of this problem, we introduce several linearization techniques to convert it into an integer linear programming (ILP) problem, which can be solved with any ILP solvers;
    
    \item We also propose a branch-and-bound algorithm (called \bnbstar) involving an A* path-finding strategy for efficient exploration of embedding possibilities. A tuning parameter is introduced in \bnbstar\ to adjust the exploration space of the algorithm, making it suitable for practical implementation when dealing with large network instances. A greedy baseline algorithm (called \bfnse) is also proposed to serve as a performance benchmark;
 
    \item Extensive numerical results for different scenarios (small-scale and large-scale networks) and settings (with and without VNF order flexibility) demonstrate that flexibility in VNF ordering improves significantly the slice acceptance rate. This enables more slices to be deployed on a constrained physical network infrastructure.
\end{itemize}

Some preliminary results on slice embedding with flexible VNF order have been presented in \cite{luu2024flexorder}. Compared to \cite{luu2024flexorder}, this paper introduces the \bnbstar\ algorithm and its variants. Extensive numerical simulations on small and large-scale networks have also been introduced.

The remainder of this paper is organized as follows. Section~\ref{sec:Related-Work} summarizes some related work. Section~\ref{sec:Problem-Statement} introduces the flexible slice embedding problem. A heuristic approach to address the embedding problem is introduced in Section~\ref{sec:Proposed-Solutions}. 
Simulation results are discussed in Section~\ref{sec:Evaluation}. Finally, Section~\ref{sec:Conclusion} concludes the work and provides some perspectives.

%%%%%%%%%%%%%%%%%%%%%%%%%%%%%%%%%%%%%
\section{Related Work}
\label{sec:Related-Work}
%%%%%%%%%%%%%%%%%%%%%%%%%%%%%%%%%%%%%

\subsection{Network Slice Embedding}

% Considering a fixed slice structure, the authors in \cite{FatemehTWC23} developed a two-timescale optimization approach that jointly optimizes traffic prediction, flow-split distribution, dynamic user association, and radio resource management to adapt to varying network conditions across time slots.
% A dynamic resource allocation framework for RAN slicing was proposed in \cite{feng2020dynamic} to manage hybrid uRLLC and eMBB services. This approach, inspired by a two-timescale algorithm with Lyapunov optimization, minimizes the total cost in terms of power consumption and negatives network utility, while satisfying strict latency requirements. However, these works primarily rely on fixed slices with a predefined VNF chain order for different services, which limits the flexibility, efficiency, and scalability of Open RAN network slices, ultimately reducing the networks ability to meet diverse QoS requirements effectively.

Most existing research on the NSE problem assumes that the order of VNFs within a given slice is predetermined \cite{qu2019delay,luo2020effective,khosravian2020ietf,cai2023privacy}. These studies typically follow a static sequence of VNFs. For example, \cite{qu2019delay} addressed the challenge of meeting end-to-end (E2E) delay requirements for embedded services in 5G networks by dynamically migrating flows to balance load and minimize reconfiguration overhead. The authors proposed a mixed-integer quadratically constrained programming (MIQCP) model to optimize this balance, considering constraints like processing capacity, resource allocation, and E2E delay. This study centers on flow migration within a fixed VNF embedding, rather than a flexible, dynamic reordering of VNFs within network slices.

In \cite{luo2020effective}, a method was presented for parallelizing and deploying service function chains (SFCs) to reduce latency in Network Function Virtualization (NFV) environments. This approach leverages dependency-based parallelism, deploying SFCs as directed acyclic graphs rather than traditional linear chains. While this reduces latency, it does not address flexible VNF composition within a single service chain. Additionally, \cite{khosravian2020ietf} introduced a finite automaton (FA) model for composing SFCs based on specific IETF scenarios, including data centers, mobile networks, and security applications. 
This model aims to lower the computational complexity of SFC validation by restricting the solution space with predefined rules, which facilitates SFC management and efficient network function operation without allowing flexible VNF order.	 

The study in \cite{cai2023privacy} focused on the privacy-preserving deployment of SFCs across multiple domains, using a deep Q-network (DQN)-based strategy. This approach enables service composition across privacy-sensitive domains without disclosing resource or topology information. However, it does not address the flexible reordering of VNFs within a single domain to optimize resource utilization.

Overall, most research on the NSE problem assumes that the order of VNFs within a given slice is fixed in advance, following a predetermined sequence that overlooks the potential benefits and challenges of flexible VNF ordering. 
This creates a notable gap in the literature around ordering VNF flexibly to form a slice before the embedding, implying the need for innovative solutions that address the advantages as well as complexities of flexible VNF ordering in network slicing.

\subsection{Network Slicing With Flexible VNF Order}

A few works address the flexibility of VNF order and the composition of network slices can be found in the literature.
For instance, in \cite{mehraghdam2016placement}, a YANG data model was proposed to support the flexibility in ordering VNFs within a network service, \emph{i.e.}, the order of some VNFs can be swapped without affecting the overall functionality of the slice. In the proposed system, the network orchestration system can select the ideal combination of service components to achieve the most effective service placement within the network. Nevertheless, this work has not provided an optimization framework to address the NSE problem with flexible VNF order.

The problem of VNF composition has been studied in \cite{ocampo2017optimal}. The paper has addressed the optimal composition of SFCs by characterizing service requests in terms of VNFs and formulating the composition as an ILP problem. This approach generates optimal SFCs by considering dependencies between VNFs and determining the best configuration to minimize resource usage, particularly bandwidth. The ILP-based solution ensures optimality, contrasting with existing heuristic approaches that may not yield the best possible configurations. Nevertheless, while this paper does explore flexible VNF composition within the context of SFCs, it primarily focuses on the SFC composition stage rather than the complete embedding process. It does not specifically address the joint problem of embedding VNFs onto a physical network infrastructure, which is critical in network slicing.

The VNF composition problem 
has also been tackled in \cite{gilherrera2017scalable}. The authors have proposed a tabu search metaheuristic (TS-SFCC) to compose SFCs, focusing on minimizing total bandwidth usage across the chain. By allowing flexibility in VNF arrangement within the chain, this approach aims to identify efficient SFCs based on the data processing and resource requirements of VNFs. The TS-SFCC algorithm prioritizes configurations that reduce bandwidth needs while ensuring scalability, making it feasible for large-scale networks. The algorithm demonstrates advantages over exact methods, especially in scenarios with complex VNF dependencies.
Similar to \cite{ocampo2017optimal}, this paper has not extended to the complete VNE process, as the focus is solely on the SFC composition without embedding VNFs onto a physical network.

The work in \cite{luu2024flexorder} is the first attempt to mathematically formulate the problem of network slice embedding with flexible VNF ordering. Nevertheless, it primarily illustrates the advantages of flexible VNF ordering through basic simulation scenarios and does not propose a suboptimal algorithm to efficiently solve the problem.

\subsection{Network Slice Embedding based on BnB and A*}

The branch-and-bound (BnB) algorithmic has widely been used to obtain exact solutions for a variety of optimization problems \cite{lawler1966branch,morrison2016branch}. With BnB, a tree search is employed, which implicitly considers all potential solutions to the problem, using pruning rules to discard sections (\textit{i.e.}, branches) of the search space that cannot yield an improved solution.

A branch-and-price optimization framework for the VNE problem has been proposed in \cite{wang2016branch}. The framework uses a path-based ILP model combined with column generation and branch-and-bound techniques. 
The authors have addressed computational efficiency by employing column generation to limit the path space and obtain near-optimal solutions with instance-specific performance guarantees. 
They focus on optimizing node and link resource allocation on the network infrastructure, reducing both the computational complexity and resource costs in the embedding process.

The work in \cite{taghavian2023cnsm} addressed the problem of the online placement of services by proposing the basics of a BnB-based approach, allowing the application of different AI search strategies. It also demonstrated how to obtain optimal results (the distinguishing advantage of ILP-based placement methods) while maintaining the scalability of a heuristic-grade placement strategy. In the extended work \cite{taghavian2023approach}, the authors propose a hybrid method combining ILP, column generation (CG), and a BnB structure integrated with AI search strategies, particularly A*. This approach aims to balance optimal service placement and scalability in large networks. The paper introduces various strategies like A*-based optimization, fairness in placement, and a parallel strategy for integrating multiple algorithms, improving service acceptance. Through extensive evaluations, the proposed solution shows significant scalability and near-optimal results, achieving a closer performance to the optimum while maintaining low execution times. This method offers a viable and efficient solution for NFV in dynamic, resource-constrained edge and cloud environments.

% \subsection{A*-based Network Slice Embedding}

The challenges of network slice placement and chaining in 5G networks have been considered in \cite{aklamanu2019utility}. The authors propose UA* (utility-based A* algorithm), an approach that integrates the traditional A* algorithm with a data-driven utility function for ranking and prioritizing VNF placement while classifying nodes into different categories. This classification system allows for more nuanced decision-making in resource allocation. The algorithm considers both computational resources (CPU, RAM, storage) and network requirements (bandwidth, latency), while also accounting for end-to-end network slice SLAs and non-linear network slice graphs.

%%%%%%%%%%%%%%%%%%%%%%%%%%%%%%%%%%%%%
\section{Problem Statement} 
\label{sec:Problem-Statement}
%%%%%%%%%%%%%%%%%%%%%%%%%%%%%%%%%%%%%

\begin{figure*}[htb]
    \centering
    \includegraphics[width=0.95\textwidth]{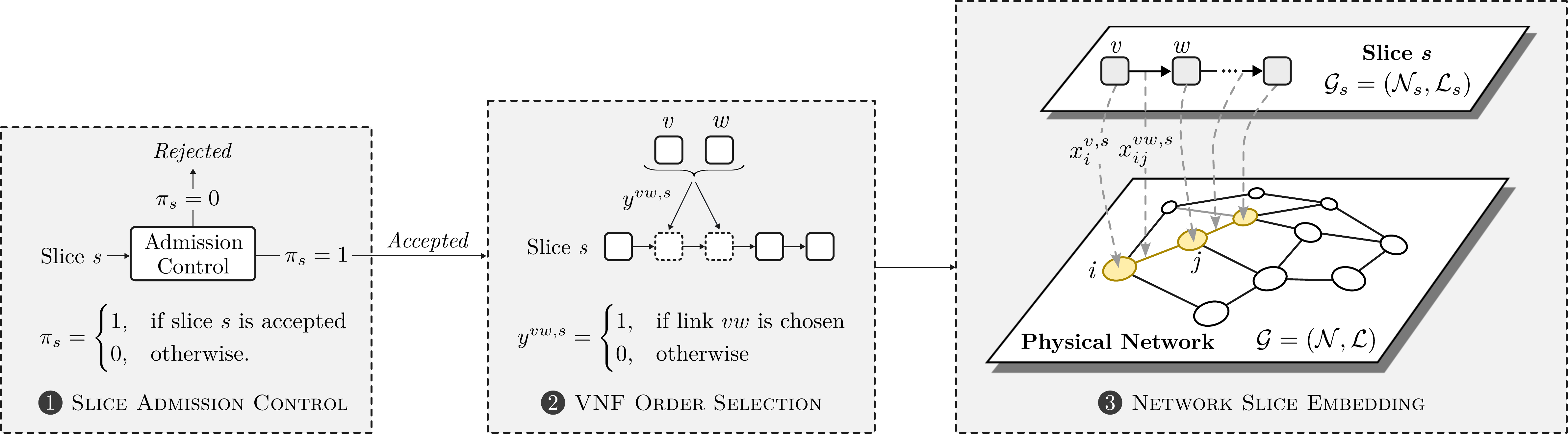}
    \caption{Illustration of the considered problem: The three steps: ($1$) slice admission control, ($2$) VNF order selection and ($3$) network slice embedding are jointly optimized. This illustration considers one single slice $s$. \label{fig:intro-problem-SEFlexOrder}}
\end{figure*}

In this section, we formalize the problem of admission control and slice embedding when flexibility in the order of their VNFs is allowed.
Upon receiving the request for deploying a given slice, the VNO can then choose the order of VNFs to form the slice that provides the best embedding solution, \emph{e.g.}, in terms of deployment cost.

%%%%%%%%%%%%%%%%%%%%%%%%%%%%%%%%%%%%%
\subsection{Network Model}
\label{sec:Network-Model}
%%%%%%%%%%%%%%%%%%%%%%%%%%%%%%%%%%%%%
The physical network infrastructure is represented as a weighted graph $\cG = (\cN, \cL)$, where $\cN$ is the set of physical nodes and $\cL$ is the set of all available links within $\cG$. Each physical node $i \in \cN$ is characterized by its available resource capacity $A_{i}$ (\emph{e.g.}, computing, storage, or processing). Similarly, each physical link $ij \in \cL$ has an available resource capacity $A_{ij}$ (\emph{e.g.}, bandwidth).

%%%%%%%%%%%%%%%%%%%%%%%%%%%%%%%%%%%%%
\subsection{Slice Model}
\label{sec:Slice-Model}
%%%%%%%%%%%%%%%%%%%%%%%%%%%%%%%%%%%%%

For each slice $s \in \cS$, let $\cN_s$ be the set of virtual nodes (representing the VNFs). Each virtual node $v \in \cN_s$ requires an amount $R_{v}$ of physical resources. 
The set 
\begin{equation}\label{FO:Define-Ps}
\cP_{s} = \{(v,w), v\in \cN_{s}, w \in \cN_{s}: w \neq v \}
\end{equation}
represents all possible combinations of VNF pairs that may be connected by a virtual link.

To simplify presentation, we consider slices represented by linear chains, without loops or branches. We assume, moreover, that some virtual nodes $v \in \cN_s$ have a fixed position, while others may be placed flexibly in the slice. In what follows, a slice \emph{configuration} represents a possible organization of the VNFs within that slice.
In the example shown in Fig.~\ref{fig:flexible-example}, 
two VNFs, VOC and TM, can be flexibly swapped with each other, creating two possible slice configurations, $(\text{IDPS} \rightarrow \text{VOC} \rightarrow \text{TM} \rightarrow \text{GW} \rightarrow \text{DU})$ and $(\text{IDPS} \rightarrow \text{TM} \rightarrow \text{VOC} \rightarrow \text{GW} \rightarrow \text{DU})$.

In general, different slice configurations can result in varying resource requirements for both the VNFs and the virtual links connecting them. For simplicity, we assume that all possible slice configurations share the same VNF resource requirements, with differences limited only to the resource demands of the virtual links.

%%%%%%%%%%%%%%%%%%%%%%%%%%%%%%%%%%%%%
\subsection{Variables}
\label{sec:Variables}
%%%%%%%%%%%%%%%%%%%%%%%%%%%%%%%%%%%%%

The general decision process for any slice $s \in \cS$ is illustrated in Fig.~\ref{fig:intro-problem-SEFlexOrder}. Briefly, upon receiving the request for a given slice $s$, the VNO has to make the following decisions:
\begin{enumerate}
    \item[($1$)] \textbf{Slice admission control:} Whether the requested slice is accepted or not;
    \item[($2$)] \textbf{VNF order selection:} If accepted, which configuration should be used to provide the requested service in a way that achieves better resource utilization; and
    \item[($3$)] \textbf{Network slice embedding:} What are the locations of physical nodes and links to embed (map) the VNFs and virtual links of the accepted slice with the chosen configuration?
\end{enumerate}

For the first decision process, one introduces the set of binary variables
$\boldsymbol{\pi} = \{\pi_{s}\}_{s \in \cS}$
to represent the slice admission control indicator, \textit{i.e.},
\begin{equation}\label{FO:Define_pi}
    \pi_{s} = 
    \begin{cases}
        1, & \text{if slice $s$ is accepted} \\
        0, & \text{otherwise}.
    \end{cases}
\end{equation}
% $\pi_{s} = 1$ if slice $s$ is accepted and $\pi_{s} = 0$, otherwise.

For the second decision process, the set of binary variables $\boldsymbol{y} = \{y^{vw,s}\}_{s \in \cS, vw \in \cP_{s}}$ is used, where $y^{vw,s}$ indicates whether or not a virtual link is formed between VNFs $v$ and $w$ to compose slice $s$, \emph{i.e.}, 
\begin{equation}\label{FO:Define_y}
    y^{vw,s}=
    \begin{cases}
        1, & \text{if virtual link }vw\text{ is chosen} \\
        0, & \text{otherwise}.
    \end{cases}
\end{equation}
The set of all virtual links\footnote{Note that $\cL_{s}$ is not known in advance. This set is only determined after a successful mapping of slice $s$ onto the physical network.} used in an slice configuration can be represented by $\cL_{s} = \{vw \in \cP_{s}: y^{vw,s} = 1\}$. 
One also introduces the variable set $\boldsymbol{\theta} = \{\theta_{p}^{v,s}\}_{s \in \cS, v \in \cN_{s}, p \in [1,\vert \cN_{s} \vert]}$, where $\theta_{p}^{v,s}$ represents the position of each VNF in slice $s$,
\begin{equation}\label{FO:Define_theta}
    \theta_{p}^{v,s} =
    \begin{cases}
        1, & \text{if } v \text{ is placed at the } p^\text{th} \text{ position of slice }s \\
        0, & \text{otherwise}.
    \end{cases}
\end{equation}
The position of VNF $v$ in slice $s$ is then
\begin{equation} \label{FO:Define_Position}
    p^{v,s}=\sum_{p=1}^{\left|\cN_{s}\right|}p\theta_{p}^{v,s},\forall v\in\cN_{s}.
\end{equation}

Now, for the final decision process, one denotes by $\boldsymbol{x} = \{x_{i}^{v, s}, x_{ij}^{vw, s}\}_{(i, ij) \in \cG, (v, vw) \in \cG_s, s \in \cS}$ the set of mapping decisions.
Here, the node mapping indicator variables $x_{i}^{v, s} \in \{0,1\}$ indicates whether or not the VNF $v$ of slice $s$ is mapped onto the physical node $i\in \cN$. Similarly, the link mapping indicator $x_{ij}^{vw, s} \in \{0,1\}$ indicates whether or not the virtual link $vw$ of slice $s$ is mapped onto the physical link $ij \in \cL$.

Table~\ref{tab:Notations} summarizes the main notations used throughout this paper.
\begin{table}[t]
\caption{Main notations. \label{tab:Notations}}
% \footnotesize
\centering
\vspace{-0.2cm}
\begin{tabular}{cl}
\toprule 
    {\textit{Symbol}} & {\textit{Description}} \tabularnewline
    \cmidrule[0.4pt](lr{0.12em}){1-1}%
    \cmidrule[0.4pt](lr{0.12em}){2-2}%
    $\mathcal{G}$& Graph representing the physical network \tabularnewline
    $\mathcal{N}$& Set of physical nodes in $\cG$\tabularnewline
    $\mathcal{L}$& Set of physical links in $\cG$\tabularnewline
    $A_{i}$& Available resources of physical node $(i)$ \tabularnewline
    $A_{ij}$& Available resources of physical link $ij$ \tabularnewline
    $\mathcal{G}_{s}$& Graph representing slice $s$ \tabularnewline
    $\mathcal{N}_{s}$& Set of VNFs in slice $s$ \tabularnewline
    $\cN_{s}^{\text{F}}$& Set of VNFs of slice $s$ whose positions are fixed\tabularnewline
    $\mathcal{L}_{s}$& Set of virtual links in slice $s$ \tabularnewline
    $\cP_{s}$& Set of all possible combinations of VNF pairs of slice $s$ \tabularnewline
    $\cP_{s}^{\text{F}}$& Set of VNF pairs of slice $s$ that are fixed to form virtual links\tabularnewline  
    $R_{v}$& Resources requirements of VNF $v$ \tabularnewline
    $R_{vw}$& Resources requirements of virtual link $vw$ \tabularnewline
    $\cS$ & Set of slices $s$\tabularnewline
    $\pi_s$& Admission control variable of slice $s$ \tabularnewline
    $x_{i}^{v, s}$& Node mapping indicator between $v$ and $i$ \tabularnewline
    $x_{ij}^{vw, s}$& Link mapping indicator between $vw$ and $ij$ \tabularnewline
    $y^{vw, s}$ &  Link combination selection variable \tabularnewline
    $\theta_{p}^{v, s}$ &  VNF position selection variable \tabularnewline
    %  $\Delta_{i}$& Cost of mapping onto node $i$ of the physical network \tabularnewline
    % $\Delta_{ij}$& Cost of mapping onto link $ij$ of the physical network \tabularnewline
\bottomrule
\end{tabular}
\end{table}

%%%%%%%%%%%%%%%%%%%%%%%%%%%%%%%%%%%%%
\subsection{Constraints}
%%%%%%%%%%%%%%%%%%%%%%%%%%%%%%%%%%%%%

In what follows, constraints are introduced to account for the flexibility in the positioning of certain VNFs within slice configurations and outline the problem of admission control and embedding of slices with flexible-order VNFs.

\noindent{\em Node capacity:}
The total requested resources in the physical node $i$ must not exceed its available resources $A_{i}$:
\begin{equation} \label{FO:C1}
\sum_{s \in \cS}\sum_{v\in\cN_{s}}x_{i}^{v,s} R_{v} \le A_{i},  \qquad\forall i\in\mathcal{N}. 
\end{equation}

\vspace{0.1cm}
\noindent{\em Link capacity:}
The total requested resources in a physical link should never exceed its available resources,
\begin{equation} \label{FO:C2-Nonlinear}
\sum_{s \in \cS}\sum_{(v,w) \in\cP_{s}}x_{ij}^{vw, s}y^{vw, s} R_{vw}\le A_{ij}, \quad \forall ij\in\mathcal{L}.
\end{equation}
The quadratic term $x_{ij}^{vw, s}y^{vw, s}$ makes  \eqref{FO:C2-Nonlinear} nonlinear. To linearize \eqref{FO:C2-Nonlinear}, one replaces $x_{ij}^{vw, s}y^{vw, s}$ by an additional variable $z_{ij}^{vw,s} \in \{0,1\}$ as 
\begin{equation}
    \sum_{s \in \cS}\sum_{(v,w) \in\cP_{s}} z_{ij}^{vw,s} R_{vw} \le A_{ij}, \quad \forall ij \in \cL
\end{equation}
and imposes the following constraints to linearize it for all $ij\in\mathcal{E}$ for  $\forall s \in \cS$ and  $\forall\left(v,w\right)\in\cP_{s}$:
\begingroup
\allowdisplaybreaks
\begin{subequations} \label{FO:C2-Linear}
\begin{align} 
z_{ij}^{vw,s} & \le x_{ij}^{vw,s} \\
z_{ij}^{vw,s} & \le y^{vw,s} \\
z_{ij}^{vw,s} & \geq x_{ij}^{vw,s}+y^{vw,s} - 1.
% & \quad\quad \forall ij\in\mathcal{E},s \in \cS,\left(v,w\right)\in\cP_{s}. 
\end{align}
\end{subequations}
\endgroup

\vspace{0.1cm}
\noindent{\em Mapped only once:}
As in \cite{Riggio2016} or \cite{luu2018aggregated}, we introduce the following constraint 
\begin{equation} \label{FO:C3}
\sum_{v\in\cN_{s}}x_{i}^{v,s} \le \pi_{s}\qquad\forall i\in\mathcal{N},s \in \cS
\end{equation}
to ensure that each physical node hosts at most one VNF per slice. This limitation helps guarantee that, in the event of a node failure, only a single function within the chain needs relocation.

\vspace{0.1cm}
\noindent{\em Accepted slices are served:}
This constraint guarantees that once slice $s$ is accepted (\emph{i.e.}, $\pi_{s} = 1$), all of its VNFs should be mapped onto the physical network:
\begin{equation}
\sum_{i\in\mathcal{N}}x_{i}^{v,s} = \pi_{s}\qquad\forall v\in\cN_{s},s \in \cS. \label{FO:C4}
\end{equation}

\vspace{0.1cm}
\noindent{\em Flow conservation:}
This constraint enforces that, once a virtual link $vw$ belongs to the chosen configuration for slice $s$, \emph{i.e.}, $y^{vw, s} = 1$, the mapping of $vw$ onto one or several physical links should preserve the traffic between VNFs $v$ and $w$,
\begin{align} \label{FO:C5-Nonlinear}
 & {\displaystyle \sum_{j\in\mathcal{N}}x_{ij}^{vw,s}-\sum_{j\in\mathcal{N}}x_{ji}^{vw,s}=x_{i}^{v,s}-x_{i}^{w,s},}\nonumber \\
 & \quad\quad\forall i\in\mathcal{N},s\in\mathcal{S},\left(v,w\right)\in\mathcal{P}_{s}:y^{vw,s}=1.
\end{align}
Constraint \eqref{FO:C5-Nonlinear} has only to be active when $y^{vw,s} = 1$. To address this issue, the big-$M$ method is considered as follows
\begin{align} \label{FO:C5-Linear}
 & \sum_{j\in\mathcal{N}}x_{ij}^{vw, s}-\sum_{j\in\mathcal{N}}x_{ji}^{vw,s} - x_{i}^{v,s} + x_{i}^{w,s}\le M\left(1-y^{vw, s}\right) \nonumber \\
 & \sum_{j\in\mathcal{N}}x_{ij}^{vw, s}-\sum_{j\in\mathcal{N}}x_{ji}^{vw,s}-x_{i}^{v,s}+x_{i}^{w,s}\geq-M\left(1-y^{vw, s}\right) \nonumber \\
 & \quad\quad \forall i\in\mathcal{N},s \in \cS,\left(v,w\right)\in\cP_{s}. 
\end{align}

\begin{figure}[tb]
    \centering
    \subfloat[\label{fig:loop-prevent-1}]{         
        \includegraphics[width=0.25\columnwidth]{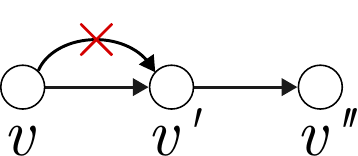}
    }\hspace{0.4cm}
    \subfloat[\label{fig:loop-prevent-2}]{      
        \includegraphics[width=0.25\columnwidth]{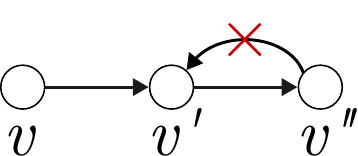}
    }\hspace{0.4cm}
    \subfloat[\label{fig:loop-prevent-3}]{
        \includegraphics[width=0.25\columnwidth]{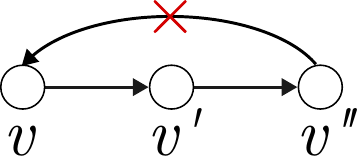}
    }
    \caption{Illustration of constraint \eqref{FO:C6c} when (a) $\mathcal{N}_{s}^{\prime} = \{v, v'\}$, (b) $\mathcal{N}_{s}^{\prime} = \{v', v''\}$, and (c) $\mathcal{N}_{s}^{\prime} = \{v, v''\}$. \label{fig:loop-prevention}}
\end{figure}

\vspace{0.1cm}
\noindent{\em Formation of virtual links:}
As only linear configurations are considered, once a given VNF pair $(v,w)$ is chosen to form the virtual link $vw$, the following constraints have to be satisfied
\begingroup 
\allowdisplaybreaks 
\begin{subequations} \label{FO:C6}
\begin{align} 
& \sum_{\left(v,w\right)\in\mathcal{P}_{s}}y^{vw,s}\leq\pi_{s},\quad\forall s\in\mathcal{S}, w \in\mathcal{N}_{s} \label{FO:C6a} \\
& \sum_{\left(v,w\right)\in\mathcal{P}_{s}}y^{vw,s}\leq\pi_{s},\quad\forall s\in\mathcal{S}, v \in\mathcal{N}_{s} \label{FO:C6b} \\
& \sum_{\left(v,w\right)\in\mathcal{N}_{s}^{\prime}}y^{vw,s}\leq\pi_{s}\left(\left|\mathcal{N}_{s}^{\prime}\right|-1\right),\quad\forall \mathcal{N}_{s}^{\prime} \subseteq\mathcal{N}_{s} \label{FO:C6c}.
\end{align}
\end{subequations}
\endgroup
In this set of constraints, 
\eqref{FO:C6a} and \eqref{FO:C6b} ensure that each VNF $v$
 has only a single incoming and outgoing virtual link, while \eqref{FO:C6c} prevents loop formation within the slice, as illustrated in Fig.~\ref{fig:loop-prevention}.

\vspace{0.1cm}
\noindent{\em Position constraint:}
Once a virtual link $vw$ is formed, \emph{i.e.}, $y^{vw,s} = 1$, the position of $v$ and $w$ has to satisfy $p^{w,s}-p^{v,s} = 1$. One has thus
\begin{equation} \label{FO:C7-Nonlinear}
p^{w,s}-p^{v,s} = 1,\quad\forall s\in\mathcal{S},\left(v,w\right)\in\mathcal{P}_{s}: y^{vw,s} = 1.  
\end{equation}
Similar to \eqref{FO:C5-Nonlinear},  constraint \eqref{FO:C7-Nonlinear} is nonlinear due to the condition $y^{vw,s} = 1$. We apply the big-$M$ method to linearize \eqref{FO:C7-Nonlinear} as
\begin{align} \label{FO:C7-Linear}
 & p^{w,s}-p^{v,s}\leq1+M\left(1-y^{vw,s}\right) \nonumber \\
 & p^{w,s}-p^{v,s}\geq1-M\left(1-y^{vw,s}\right) \nonumber \\
 & \quad\forall s\in\mathcal{S},\left(v,w\right)\in\mathcal{P}_{s}.
\end{align}

\vspace{0.1cm}
\noindent{\em Remove redundant $x_{ij}^{vw,s}$:} To ensure that if $y^{vw,s} = 0$, \emph{i.e.}, no virtual link is formed between VNFs $v$ and $w$, the following constraint is introduced
\begin{equation} \label{FO:C8-Nonlinear}
\sum_{ij\in\mathcal{L}} x_{ij}^{vw,s}=0, \quad\forall s\in\mathcal{S},\left(v,w\right)\in\mathcal{P}_{s}: y^{vw,s} = 0.
\end{equation}
Again, as in \eqref{FO:C5-Nonlinear} and \eqref{FO:C7-Nonlinear}, the big-$M$ method is used to linearize \eqref{FO:C8-Nonlinear} as follows:
\begingroup 
\allowdisplaybreaks 
\begin{subequations} \label{FO:C8-Linear}
\begin{align} 
 & \sum_{ij\in\mathcal{L}}x_{ij}^{vw,s}\leq My^{vw,s},\quad\forall s\in\mathcal{S},\left(v,w\right)\in\mathcal{P}_{s} \\
 & \sum_{ij\in\mathcal{L}}x_{ij}^{vw,s}\geq-My^{vw,s},\quad\forall s\in\mathcal{S},\left(v,w\right)\in\mathcal{P}_{s}.
\end{align}
\end{subequations}
\endgroup

\vspace{0.1cm}
\noindent{\em Only one position:} The following constraints guarantee that, once a given slice $s$ is admitted, each VNF $v$ of slice $s$ should occupy only one position $p$ in the chain (constraint~\eqref{FO:C9a}) and each position $p$ is occupied by only one VNF (constraint~\eqref{FO:C9b}):
\begingroup 
\allowdisplaybreaks 
\begin{subequations} \label{FO:C9}
\begin{align}
\sum_{p=1}^{\left|\mathcal{N}_{s}\right|}\theta_{p}^{v,s} 
&= \pi_{s},\quad\forall s\in\mathcal{S},v\in\mathcal{N}_{s}  \label{FO:C9a} \\
\sum_{v\in\mathcal{N}_{s}}\theta_{p}^{v,s} 
&= \pi_{s},\quad\forall s\in\mathcal{S},p\in\left[1,\left|\mathcal{N}_{s}\right|\right]. \label{FO:C9b}
\end{align}
\end{subequations}
\endgroup

\vspace{0.1cm}
\noindent{\em Fixed order and position constraints:}
In practice, some VNFs may occupy fixed positions within a slice. For instance, in RAN slicing, the virtual distributed unit (vDU) must be positioned at the beginning (for uplink) or the end (for downlink) of the slice \cite{polese2023understanding}. Similarly, certain VNFs must follow a specific sequence, such as in a video streaming slice where a virtual firewall is placed before the virtual traffic monitor \cite{savi2021impact}. These requirements impose constraints not only on the positions of certain VNFs but also on certain virtual links within the slice. To formalize this, let $\cP_{s}^{\text{F}} \subset \cP_{s}$ represent the set of VNF pairs with \textit{fixed} virtual links, $\cN_{s}^{\text{F}} \subset \cN_{s}$ the set of VNFs with \textit{fixed} positions, and $\cN_{s}^{\text{X}} \subset \cN_{s}$ the set of VNFs with flexible positions, \emph{i.e.}:
\begingroup 
\allowdisplaybreaks 
\begin{subequations} \label{FO:C10-org}
\begin{align}
    \cP_{s}^{\text{F}} &= \{ (v,w) \in \cP_{s}: y^{vw,s}=1\} \\
    \cN_{s}^{\text{F}} &= \{ v \in \cN_{s}: \theta_{p}^{v,s} = 1\} \\
    \cN_{s}^{\text{X}} &= \cN_{s} \setminus \cN_{s}^{\text{F}}.
\end{align}
\end{subequations}
\endgroup
One has thus
\begingroup 
\allowdisplaybreaks 
\begin{subequations} \label{FO:C10}
\begin{align}
 y^{vw,s} &= 1,\quad\forall s\in\mathcal{S},\left(v,w\right)\in\cP_{s}^{\text{F}} \\
 \theta_{p}^{v,s} &=\pi_{s},\quad\forall s\in\mathcal{S},v\in \cN_{s}^{\text{F}}.
\end{align}
\end{subequations}
\endgroup

% FINAL FORMULATION OF SE-FlexOrder

Finally, the problem of slice admission control and embedding with flexible ordered VNFs (denoted as \ref{prob:SEflexorder}), which aims to maximize the number of accepted slices and minimize the number of used physical links is described as
\begin{align} \label{prob:SEflexorder}
\underset{
    \boldsymbol{\pi}, \boldsymbol{x}, \boldsymbol{y}
}{\text{max}} \quad 
& \gamma N(\boldsymbol{\pi}) - (1-\gamma) H(\boldsymbol{x}) \tag{SE-FO}   \\
\text{s.t.} \quad 
& 
\eqref{FO:Define_Position}, 
\eqref{FO:C1}, 
\eqref{FO:C2-Linear},
\eqref{FO:C3},
\eqref{FO:C4},
\eqref{FO:C5-Linear}, 
\eqref{FO:C6}, 
\eqref{FO:C7-Linear}, 
\eqref{FO:C8-Linear}, 
\eqref{FO:C9}, 
\eqref{FO:C10}  \nonumber\\
& \boldsymbol{\pi} \in \{0,1\}, \boldsymbol{x} \in \{0,1\},  \boldsymbol{y} \in \{0,1\} \nonumber
\end{align}
% \end{subequations}
% \endgroup
where 
$N(\boldsymbol{\pi}) = \sum_{s \in \cS}\pi_s$ is the number of accepted slices,  
$H(\boldsymbol{x}) = \sum_{s\in\mathcal{S}} \sum_{vw\in\mathcal{L}_s}\sum_{ij\in\mathcal{L}}x_{ij}^{vw,s}$ 
represents the number of physical links used by slice $s$, 
and $\gamma \in [0,1]$ is a tuning parameter to balance the value between $N(\boldsymbol{\pi})$ and $H(\boldsymbol{x})$.

\begin{lemma}
There exists a polynomial time reduction of the problem of embedding network slices with fixed ordered VNFs (denoted as SE-FX) to Problem~\eqref{prob:SEflexorder}. Given that SE-FX is a well-known \textit{NP}-hard problem \cite{Riggio2016,amaldi2016computational}, it follows that Problem~\eqref{prob:SEflexorder} is also \textit{NP}-hard.
\end{lemma}

Due to its \textit{NP}-hardness, Problem~\eqref{prob:SEflexorder} is challenging to solve and becomes intractable in large-scale scenarios with extensive networks or numerous slices to embed. Therefore, in Sec.~\ref{sec:Proposed-Solutions}, we propose a low-complexity heuristic algorithm that enhances scalability and efficiently yields near-optimal solutions.

%%%%%%%%%%%%%%%%%%%%%%%%%%%%%%%%%%%%%
\section{Proposed Solutions}
\label{sec:Proposed-Solutions}
%%%%%%%%%%%%%%%%%%%%%%%%%%%%%%%%%%%%%

In this section, we present two suboptimal algorithms to solve \eqref{prob:SEflexorder}.
The first heuristic, \bnbstar\ (A*-reinforced Branch-and-Bound), combines a branch-and-bound approach and the A* path-finding algorithm. The second heuristic, \bfnse\ (Best-Fit Neighbor), is a lighter, greedy approach. Details of both \bnbstar\ and \bfnse\ are provided below.

%%%%%%%%%%%%%%%%%%%%%%%%%%%%%%%%%%%%%
\subsection{\bnbstar\ Algorithm}
%%%%%%%%%%%%%%%%%%%%%%%%%%%%%%%%%%%%%

\begin{figure}[htb]
\begin{centering}
\includegraphics[width=.9\columnwidth]{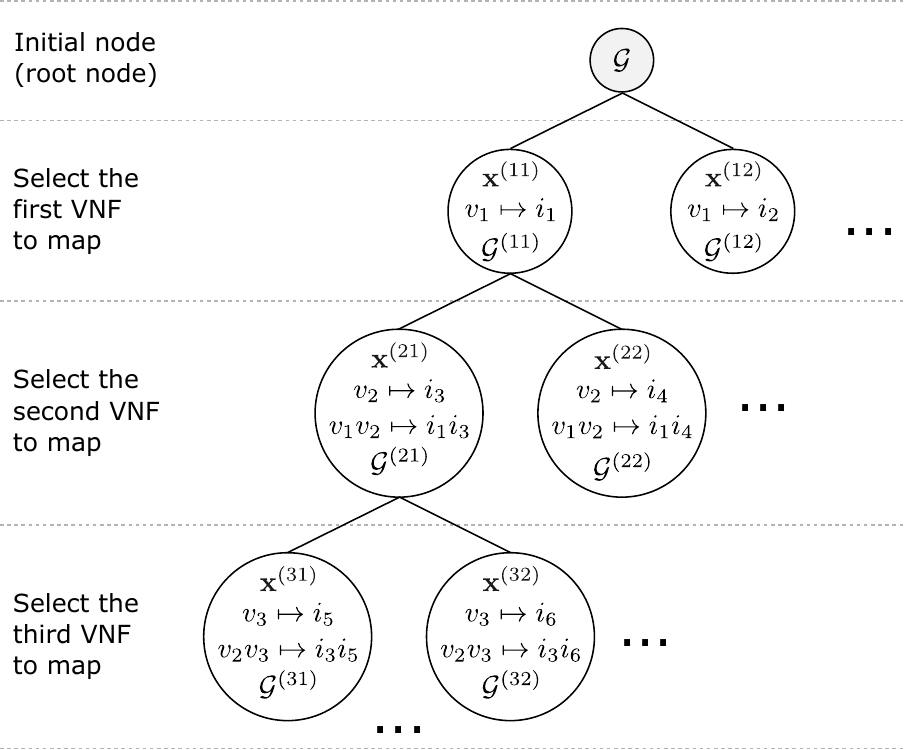}
\par\end{centering}
\caption{Illustration of the branching process in \bnbstar. Starting from the root node, where the status of the available physical resources is represented by the physical graph $\cG$. At each depth $p$ of the BnB tree, \bnbstar\ selects the candidate physical node to map the $p$-th VNF ($v_{p}$) of the considered slice. Each candidate $i_{m}$ for the mapping ($i_{m} \mapsto v_{p}$) adds a new branch to the BnB tree. The mapping choice is registered in the vector $\boldsymbol{\mathrm{x}}^{(pm)}$. The status of the physical graph after the mapping is then updated as $\cG^{(pm)}$.}
\label{fig:bnbstar-idea}
\end{figure}

% \vspace{0.1cm}
\noindent{\bf Branch-and-Bound Exploration:} 
\bnbstar\ finds feasible mapping solutions for all VNFs in a given slice
using a branch-and-bound approach.
Precisely, the branching process uses search and recursion (backtracking) to enumerate all possible mappings, as illustrated in Fig.~\ref{fig:bnbstar-idea}.
For the bounding process, inspired by the work in \cite{taghavian2023approach}, we also use the A* search algorithm to prune the branches (\textit{i.e.}, embedding possibilities) that do not provide a proper solution to Problem~\eqref{prob:SEflexorder}. This significantly reduces the computation time required to achieve the slice embedding results.

\vspace{0.1cm}
\noindent{\bf Depth First Search vs. Breadth First Search:} For a general BnB search, a depth first search (DFS) or breadth first search (BFS) strategy can be implemented, as discussed in \cite{taghavian2023approach}. 
For our problem, a DFS approach enables us to obtain an initial slice embedding solution quickly. In contrast, BFS requires more extensive exploration before reaching a feasible solution. With DFS, once an initial feasible solution with an associated cost is found, \bnbstar\ subsequently considers only those feasible solutions with a lower cost than the current best. In the case of BFS, the algorithm selects the branch with the minimum cost at each depth level, which may ultimately lead to an infeasible embedding. Therefore, we adopt only the DFS search strategy for our implementation.

\vspace{0.1cm}
\noindent{\bf Path Finding with A*:} 
A* is an efficient algorithm for finding the shortest path between two nodes of any graph. Its exploration logic is the same as Dijkstra algorithm, except for the cost function: In Dijkstra algorithm, the cost function is defined as the total cost that the path is used, while in A*, two different costs are considered: ($i$) the \textit{actual} cost $g(p)$ and ($ii$) the \textit{estimated} (or \textit{heuristic}) cost $h(p)$, where $p$ is the current node (position) in the graph, in the middle of the path searching. While the actual cost in A* is equivalent to the cumulative cost used in Dijkstra algorithm, the estimated cost provides a forward-looking estimate that helps guide the search more efficiently toward the goal.
 
In this work, we adapt the cost function for the A{*} path searching as follows 
\begin{equation}
    c_{\text{A{*}}}=g_{s}(p) + h_{s}(p) 
\end{equation}
where $g_{s}(p)$ and  $h_{s}(p)$ are respectively the actual and the estimated cost yielded by a mapping solution of slice $s$ at the $p$-th depth of the BnB exploration tree.

We define the actual cost $g_{s}(p)$ as the sum of normalized node and link resources usage by the solution at the $p$-th depth in the BnB tree of slice $s$:
\begin{align} \label{eq:g-func}
    g_{s}(p) = \rho_{1} \sum_{i\in\mathcal{N}} \sum_{v\in\mathcal{N}_{s}} \frac{R_{i}}{A_{i}} x_{i}^{v,s} + \rho_{2} \sum_{ij\in\mathcal{L}} \sum_{vw\in\mathcal{L}_{s}} \frac{R_{vw}}{A_{ij}} x_{ij}^{vw,s} 
\end{align}
where $\rho_{1}$ and $\rho_{2}$ indicate the weight associated with the usage of physical nodes and links, respectively.

For the estimated cost $h_{s}(p)$, we aim to balance the distribution of slice resource requirements to facilitate easier embedding of subsequent slices. Then
\begin{equation} \label{eq:h-func}
    h_{s}(p) = 
    \rho_{1} \frac{\sigma(\{A_{i} (p) \}_{i \in \cN})}{\sum_{i\in\mathcal{N}}A_{i}} 
    + \rho_{2} \frac{\sigma(\{A_{ij} (p) \}_{ij \in \cL})}{\sum_{ij\in\mathcal{L}}A_{ij}}
\end{equation}
where $\sigma(\{A_{i} (p) \}_{i \in \cN})$ and $\sigma(\{A_{ij} (p) \}_{ij \in \cL})$ are the standard deviation of the remaining resources of all physical nodes $i \in \cN$ and links $ij \in \cL$, respectively,  corresponding to the current solution at depth $p$ in the BnB tree of slice $s$. The two terms in \eqref{eq:h-func} are weighted normalized standard deviations, with weights $\rho_{1}$ and $\rho_{2}$, as in \eqref{eq:g-func}.

\vspace{0.1cm}
\noindent{\bf Detailed Workflow of \bnbstar:} 
The pseudocode of \bnbstar\ is summarized in Algorithms~\ref{algo:BnBStar} ($\textsc{Main}$ function) and \ref{algo:Backtrack} ($\textsc{Backtrack}$ function).

\noindent{\underline{\textsc{Main}}:}
\bnbstar\ works on a sequential manner, \textit{i.e.}, slices are embedded sequentially onto the physical network $\cG$. For each slice $s \in \cS$, the algorithm generates the set $\cK_{s}$ of all possible configurations that slice $s$ may have (Algo.~\ref{algo:BnBStar}, Line~$3$). 
Then, it loops through each configuration $k \in \cK_{s}$ of slice $s$ and performs the following steps.

For each configuration,  the algorithm resets the cost $c = c_{\text{min}}$, initializes the current solution $x_{\text{curr}}$, and calls the \textsc{Backtrack} function to perform the brand-and-bound exploration (Lines $8$--$15$).
If a feasible solution is found and the cost is lower than the current minimum cost ($c < c_{\text{min}}$), it updates the best solution $x_s = x_{\text{curr}}$ and the minimum cost (Lines $16$--$18$). If no feasible solution is found, \bnbstar\ proceeds to the next configuration $k \in \cK_{s}$.

Finally, if the set of feasible solutions of slice $s$ ($\boldsymbol{x}_{s}$) is not empty, then slice $s$ is accepted and it will be embedded into the physical network using the selected configuration $k$ associated to the solution. Otherwise, slice $s$ is rejected, and the algorithm continues with the next slice (Lines $21$--$27$).

\noindent{\underline{\textsc{Backtrack}}:}
This function attempts to map the VNFs of the slice onto physical nodes. 
If all VNFs have been mapped (\textit{i.e.}, a feasible solution has been found), the function stops performing backtracking for this solution. Otherwise, it proceeds with the current VNF $v = \mathcal{N}_s[p]$ and explores its mapping options (Lines $2$--$5$).

In the main body of $\textsc{Backtrack}$, a resource feasibility check is executed to verify whether the selected physical node $i$ has sufficient resources to host the VNF (Lines $15$--$16$). If resources are insufficient, the function skips this node. Then, if a previous VNF has successfully been mapped onto physical node $j$, it attempts to find the shortest path between the current node $i$ and the previously mapped node $j$. If no path exists, the function skips this configuration (Lines $18$--$21$). Otherwise, if a feasible path is found, the cost is updated using the A* algorithm, and the current solution is revised to include the new node and link mappings (Lines $22$--$24$).

The function then updates the status of the mapping of VNF $v$; if it has been successfully mapped, the algorithm proceeds to the next VNF $p + 1$. If no feasible solution is found, the function returns with an ``infeasible'' flag (Lines $31$--$35$).

$\textsc{Backtrack}$ is called recursively to explore slice embedding solutions that may yield a lower cost than the current one. It terminates only after all branches of the branch-and-bound tree have been fully explored. 

% MAIN OF BNBSTAR
\begin{algorithm}[t]
\footnotesize
\caption{\bnbstar\ algorithm. \label{algo:BnBStar}}
\BlankLine
\everypar={\n}
\KwInput{$\mathcal{G}, \mathcal{S}$}
\KwOutput{$\boldsymbol{\mathrm{x}} = 
\{
    \langle
        \pi_{s}, \boldsymbol{x}_{s}, k_{s}
    \rangle \vert s \in \cS
\}$}
\SetKw{return}{return}
\SetKw{continue}{continue}
\SetKw{break}{break}
\SetKw{true}{True}
\SetKw{and}{and}
\SetKw{false}{False}
\SetKw{or}{or}
\SetKw{init}{Initialize:}
% \SetKw{func}{Function}
\SetKwProg{Fn}{Function}{:}{end}
\BlankLine

\Fn{\textsc{Main}}{
    \ForEach{slice $s\in\mathcal{S}$}{
		$\cK_{s} = \textsc{GetAllPossibleConfigs}(s)$;
  
		$\boldsymbol{x}_{s} = \varnothing$; \Comment{Initial embedding solution for slice $s$}
		
		$c_{\text{min}} = \infty$; \Comment{Initial minimum cost}

        $k_{s} = \varnothing$; 
        \Comment{Initial config for slice $s$}
		
		\ForEach{configuration $k \in \cK_{s}$}{

			$\cG_{s} = \cG_{s}^{k}$; 
                \Comment{Take the graph of config $k$}

			$c = c_{\min}$;
                \Comment{Initial temporary cost}
			
			$\boldsymbol{x}_{\text{curr}} = \varnothing$;
                \Comment{Initial current solution}
			
			$p = 0$;
                \Comment{Index of the current VNF to map}

			\text{flag} $=$ ``infeasible'';
                \Comment{Initial feasibility flag}

                \vspace{3pt}
                \textit{\# Update $c, \boldsymbol{x}_{\text{curr}}, p, \text{flag}$ using \textsc{Backtrack}}

            $\beta=\infty$ \Comment{Disable branch limit.}
                
			$\textsc{Backtrack}(\cG, \cG_{s}, c,
                \boldsymbol{x}_{\text{curr}}, p, \text{flag}, \beta)$;
                \vspace{3pt}

			\eIf{(flag $==$ ``feasible'') \and ($c < c_{\text{min}}$)}{

				$\boldsymbol{x}_{s} = \boldsymbol{x}_{\text{curr}}$; \Comment{Update solution}

				$c_{\text{min}} = c$; 
                \Comment{Update new minimum cost}
			}{
				\continue; 
                \Comment{Try with next config}
			}
		} % EndFor

		\eIf{$\boldsymbol{x}_{s} \neq \varnothing$}{
			$\pi_{s} = 1$; 
                \Comment{Accept slice $s$}

			Map slice $s$ with selected configuration $k$;

                $k_{s} = k$; 
                \Comment{Update final config for slice $s$}
		}{
			$\pi_{s}=0$; 
                \Comment{Reject slice $s$}
			
			\continue; 
                \Comment{Continue with next slice}
		} % EndIf
	} % EndFor
} % EndFunction

\end{algorithm}

% FUNCTION BACKTRACK
\begin{algorithm}[htb]
\footnotesize
\caption{Function \textsc{Backtrack} of \bnbstar. \label{algo:Backtrack}}
\BlankLine
\SetKw{return}{return}
\SetKw{continue}{continue}
\SetKw{break}{break}
\SetKw{true}{True}
\SetKw{and}{and}
\SetKw{false}{False}
\SetKw{or}{or}
\SetKw{init}{Initialize:}
% \SetKw{func}{Function}
\SetKwProg{Fn}{Function}{:}{end}
% \BlankLine

\Fn{$\textsc{Backtrack}(\cG, \cG_{s}, c, \boldsymbol{x}_{\text{curr}}, p, \text{flag},\beta)$}{

    \If{$p \geq \left| \cN_{s} \right|$}{
        \text{flag} $=$ ``feasible''; \Comment{All VNFs have been mapped}

        \return;
    }

    $v=\cN_{s}[p]$; 
    \Comment{Get the current VNF}

    \If{$p > 0$}{
        $w=\cN_{s}[p-1]$; 
        \Comment{Get the previous VNF}

        Get $j\in\cN_{s}$ s.t. $x_{j}^{w,s}=1$; 
        \Comment{Get previous physical node}
    }

    $n_{\text{sol}}=0$; \Comment{Initialize solution counter}
    
    \ForEach{$i\in\mathcal{N}$}{    \Comment{Loop on physical nodes}
    
        \If{$n_{\mathrm{sol}}\geq \beta$}{
        \return; \Comment{Solution limit reached, stop exploring}
        }
        
        \If{$\sum_{v'\in\cN_{s}:v'\neq v} x_{i}^{v',s} = 1$}{
            \continue; 
            \Comment{Node $i$ has already been used, skip}
        }
        \If{$a_i < r_v$}{
            \continue; 
            \Comment{Lack of resources in node $i$, skip}
        }

        $x_{i}^{v,s}=1$,      
        \Comment{Update node mapping result}

        \If{$p > 0$}{
            $P^{\star}_{ji} = \textsc{ShortestPath}(j,i)$;
            \Comment{Find shortest path from $j$ to $i$}
        
            \If{$P^{\star}_{ji} = \varnothing$}{
                \continue; 
                \Comment{No path available from $j$ to $i$, skip}
            }

            $x_{\ell}^{wv,s}=1,  \forall \ell \in P^{\star}_{ji} $;
            \Comment{Update link mapping result}
        }

        $c_{A^{\star}} = g_{s} + h_{s}$;
        \Comment{Calculate $A^{\star}$ cost}
        
        $c_{\mathrm{curr}} = c + c_{A^{\star}}$;
        \Comment{Update current cost}

        \eIf{$c_{\mathrm{curr}} < c$}{
            $c = c_{\mathrm{curr}}$; 
            \Comment{Update temporary cost $c$}
            
            $\boldsymbol{x}_{\mathrm{curr}} =
            \{ x_{i}^{v,s}, x_{\ell}^{wv,s} \}$;
            \Comment{Update current solution}
        }{
            \continue; \Comment{Prune this branch}
        }

        \textit{\# Finally, if VNF $v$ is successfully mapped}
        
        \eIf{$\sum_{i\in\mathcal{N}} x_{i}^{v,s} = 1$}{
            $p = p + 1$; \Comment{Go to next VNF}
        }{
            \text{flag} $=$ ``infeasible''; 
            \Comment{VNF $v$ cannot be mapped, stop}
            
            \return;
        } % EndIf
    
        $\textsc{Backtrack}(\cG, \cG_{s}, c, \boldsymbol{x}_{\text{curr}}, p, \text{flag})$; \hspace{-0.4cm}\Comment{Recall \textsc{Backtrack}}

        $n_{\text{sol}}=n_{\text{sol}}+1$; \Comment{Solution found}
    } % EndFor
	
} % EndFunction

\end{algorithm}

\vspace{0.1cm}
\noindent{\bf Variants of \bnbstar:}  A threshold 
$\beta$ on the number of explored solutions can be set (Line $11$).
By adjusting $\beta$ (\textit{i.e.}, terminating \bnbstar\ when reaching exactly $\beta$ feasible solutions), one can control the trade-off between computation time and the quality of the slice embedding solution. The more solutions found, the higher the slice acceptance rate, but this also increases the computational cost. The resulting algorithm is called  \bnbstar-$\beta$ in what follows. Then \bnbstarinf\ corresponds to $\beta = \infty$. Without further mention, \bnbstar\ refers to \bnbstarinf.

\vspace{0.1cm}
\noindent{\bf Complexity Analysis of \bnbstar:} 
First, at every node of the BnB tree, in the worst case scenario, the \textsc{ShortestPath} function (Algo.~\ref{algo:Backtrack}, Line~19) is run. Its complexity is that of Dijkstra algorithm, \emph{i.e.}, $\mathcal{O}((\vert \mathcal{N} \vert + \vert \mathcal{L} \vert) \times \log(\vert \mathcal{L}\vert))$, for every other physical nodes. Thus, the complexity of every node in the BnB tree is $\mathcal{O}((\vert \mathcal{N} \vert - 1) \times (\vert \mathcal{N} \vert + \vert \mathcal{L} \vert) \times \log(\vert \mathcal{L}\vert))$. 

Second, a depth level of the BnB tree contains $\vert \mathcal{N} \vert$ nodes, and \bnbstar\ may need to explore all of them, leading to complexity at each depth level of BnB tree of $\mathcal{O}(\vert \mathcal{N} \vert \times (\vert \mathcal{N} \vert - 1) \times (\vert \mathcal{N} \vert + \vert \mathcal{L} \vert) \times \log(\vert \mathcal{L}\vert)$).

Next, the BnB tree has a depth of at most $\vert\mathcal{N}_s\vert$. The complexity for traversing all levels of the BnB tree for a given configuration $k\in\mathcal{K}_s$ of a slice $s\in\mathcal{S}$ is approximately $\mathcal{O}(\vert \mathcal{N}_s \vert \times \vert \mathcal{N} \vert \times (\vert \mathcal{N} \vert - 1) \times (\vert \mathcal{N} \vert + \vert \mathcal{L} \vert) \times \log(\vert \mathcal{L}\vert))$, which can be simplified as $\mathcal{O}(\vert\mathcal{N}_s\vert \times \vert\mathcal{N}\vert^{3} \times \log(\vert \mathcal{L}\vert))$.

Finally, \bnbstar\ attempts to construct and traverse a BnB tree for every configuration $k\in\mathcal{K}_s$ (Algo.~\ref{algo:BnBStar}, Line~7) of each slice $s\in\mathcal{S}$ (Algo.~\ref{algo:BnBStar}, Line~2) in a sequential manner. Each permutation of flexible VNFs creates a configuration. Recall that $\mathcal{N}_{s}^{\text{X}}$ is the set of VNFs whose positions are flexible in  slice $s$, the total number of possible configuration for the slice $s$ is thus $\vert \mathcal{N}_{s}^{\text{X}}\vert!$.
Based on the above analysis, the total complexity of \bnbstar\ is
$\mathcal{O}(\vert \mathcal{S}\vert \times \vert\mathcal{N}_{s}^{\text{X}}\vert! \times \vert\mathcal{N}_s\vert \times \vert\mathcal{N}\vert^{3} \times \log(\vert \mathcal{L}\vert))$. 

In the worst case, even with a limited $\beta$, \bnbstar\ has to explore all existing branches to find $\beta$ feasible solutions. Hence, the worst-case complexity remains the same for all variants \bnbstar-$\beta$.
Nevertheless, in the average case, if the search space can be pruned effectively, the runtime may appear proportional to the value of $\beta$.

%%%%%%%%%%%%%%%%%%%%%%%%%%%%%%%%%%%%%
\subsection{\bfnse\ Algorithm}
%%%%%%%%%%%%%%%%%%%%%%%%%%%%%%%%%%%%%
To get a baseline for performance comparison with \ilpse\ and \bnbstar, we consider a simple heuristic, called \bfnse\ (Best-Fit Neighbor). \bfnse\  tries to embed each slice on the physical network in a greedy manner. Details of \bfnse\ are provided in the following.

\vspace{0.1cm}
\noindent{\bf Detailed Workflow of \bfnse:} The pseudo-code for \bfnse\ is presented in Algorithm~\ref{algo:MRN-SE}. It consists of two functions: \textsc{Main} and \textsc{MapConfig}.

For each slice $s\in\mathcal{S}$ that allows flexibility in VNF ordering, \bfnse\ determines the set $\cK_{s}$ of all possible configurations of slice $s$ (Line~$3$). Then, it tries to map each configuration $k \in \cK_{s}$ onto the physical network $\cG$ using function $\textsc{MapConfig}$ (Lines~$5$--$10$). Precisely, the first VNF $v$ of the configuration is mapped onto $\cG$ using a greedy approach, \emph{i.e.}, selecting the physical node  $i$ providing the most resource (Line~$22$). The adjacent VNF $w$ of $v$ in the configuration is then mapped onto $\cG$ by selecting the physical node $j\in\mathcal{N}$ having the most resource capacity and by prioritizing the node that is closest to the previously selected physical node $i$ (Line~$24$).  

Next, the algorithm uses Dijkstra algorithm (function \textsc{ShortestPath}) to find the shortest path $\texttt{Path}^{\star}_{ij}$ between physical nodes $i$ and $j$ to map the virtual link $vw$ between VNFs $v$ and $w$ (Lines~$25$--$31$). When all VNFs and virtual links of the current configuration $k$ of slice $s$ are mapped, the configuration mapping is considered as successful and the current configuration $k$ is added to the set $\mathcal{M}_{s}$ of \emph{mappable} configurations (Line~$8$). 

After attempting to map all configurations of slice $s$ in $\cK_{s}$ if $\mathcal{M}_{s} \neq \varnothing$, meaning that at least one configuration has been successfully mapped onto the physical network, \bfnse\  selects the configuration $k$ that provides the best objective value to problem~\eqref{prob:SEflexorder}. 
When several configurations yield the same best objective value, one configuration is chosen randomly (Lines~$12$--$14$). 
If no configuration of slice $s$ can be mapped onto the physical network, slice $s$ is rejected ($\pi_{s} = 0$) and the algorithm continues to the next slice (Lines~$17$--$19$). The pseudocode of \bfnse\ is summarized in Algorithm~\ref{algo:MRN-SE}.

\begin{algorithm}[htb]
\footnotesize
\caption{\bfnse\ algorithm. \label{algo:MRN-SE}}
\BlankLine
\everypar={\n}
\KwInput{$\mathcal{G}, \mathcal{S}$}
\KwOutput{$\langle \boldsymbol{\pi}, \boldsymbol{x}, \boldsymbol{y}, \boldsymbol{\theta}  \rangle$}
\SetKw{return}{return}
\SetKw{continue}{continue}
\SetKw{break}{break}
\SetKw{true}{True}
\SetKw{false}{False}
\SetKw{or}{or}
\SetKw{init}{Initialize:}
% \SetKw{func}{Function}
\SetKwProg{Fn}{Function}{:}{end}
\BlankLine

\Fn{\textsc{Main}}{
    \ForEach{slice $s\in\mathcal{S}$}{
    $\cK_{s} \leftarrow \textsc{GetAllPossibleConfigs}($s$)$;
    
    $\mathcal{M}_{s} = \varnothing $; \Comment{Set of mappable configurations}
    
    \ForEach{configuration $k \in \cK_{s}$}{
        $(\langle x_{i}^{v}, x_{ij}^{vw} \rangle, \texttt{isMapped}) \leftarrow  \textsc{MapConfig}(k, \cG)$;

        \eIf{\texttt{isMapped} $=$ \true}{
            $\mathcal{M}_{s} = \mathcal{M}_{s} \cup \{k\}$; \Comment{Append $k$ to $\mathcal{M}_{s}$}
        }{
            \continue; \Comment{Try with next configuration}
        }
    }
    \eIf{$\mathcal{M}_{s} \neq \varnothing $}{
        Select the best configuration $k$ from $\mathcal{M}_{s}$;
        
        \If{Multiple configuration found}{
            Randomly choose a configuration $k$;    
        }
           
        $\pi_{s}\leftarrow1$; \Comment{Accept slice $s$}
           
        Map the slice $s$ with selected configuration $k$;
    }{
        $\pi_{s}\leftarrow0$; \Comment{Reject slice $s$}
        
        \continue; \Comment{Continue with next slice}
    }

}
}

\Fn{\textsc{MapConfig}($k, \mathcal{G}$)}{

    $\texttt{infeasible}\leftarrow$\false;

    Select physical node $i \in \cN$ having most resources to map the first VNF $v$;

\ForEach{$w\in\mathcal{N}_{s}, w\neq v$}{
    Select physical node $j \in \cN$ in proximity to $i$ that has the most resources to map $w$;
    
    $\texttt{Path}^{\star}_{ij} = \textsc{ShortestPath}(i,j)$ %\Comment{Find shortest path from $i$ to $j$};
    
    \eIf{$A_{j} < R_{w}$ \or $\texttt{Path}^{\star}_{ij} = \varnothing$}{
        $\texttt{infeasible}\leftarrow$\true;

        % \textbf{break};
        \break;
    }{
        $x_{j}^{w}\leftarrow1$;
        
        $x_{ij}^{vw}\leftarrow1$, for all physical links $ij \in \texttt{Path}^{\star}_{ij}$;    
    }
}

\eIf{$\texttt{infeasible} = \textbf{True}$}{
    \return Failed result;
}{
    \return $\langle x_{i}^{v}, x_{ij}^{vw} \rangle$;
}

}
\end{algorithm}

\vspace{0.1cm}
\noindent{\bf Complexity Analysis of \bfnse:}
The complexity of Dijkstra algorithm in Line~$25$ is $\mathcal{O}((\vert \cN \vert +\vert \cL \vert) \times \log{\vert \mathcal{L}} \vert )$. The $\textsc{MapConfig}$ function tries to map every virtual function sequentially (Line~$23$), thus requiring $\mathcal{O}(\vert\mathcal{N}_{s}\vert\times( (\vert\mathcal{N}\vert+\vert\mathcal{L}\vert)\times\log{\vert\mathcal{L}}\vert)))$.
The $\textsc{Main}$ function examines every possible configuration using the $\textsc{MapConfig}$ function (Line~$6$). As mentioned above, each permutation of flexible VNFs creates a configuration. In addition, the algorithm tries to map slices iteratively (Line~$2$). In summary, the time complexity of \bfnse\ is 
$\mathcal{O}(\vert\mathcal{S}\vert \times \vert\mathcal{N}^{\text{X}}_{s}\vert! \times \vert\mathcal{N}_{s}\vert \times (\vert\mathcal{N}\vert+\vert\mathcal{L}\vert)\times\log{\vert\mathcal{L}\vert})$.

%%%%%%%%%%%%%%%%%%%%%%%%%%%%%%%%%%%%%
\section{Performance Evaluation} 
\label{sec:Evaluation}
%%%%%%%%%%%%%%%%%%%%%%%%%%%%%%%%%%%%%
In this section, we evaluate the performance of three approaches to solve problem~\eqref{prob:SEflexorder}: ($i$) \ilpse\ that provides exact and optimal solutions; ($ii$) \bnbstar-$\beta$ algorithm with $\beta=3$ and $\beta=\infty$; and ($iii$) the greedy algorithm \bfnse.

%%%%%%%%%%%%%%%%%%%%%%%%%%%%%%%%%%%%%
\subsection{Simulation Setup}
\label{subsec:simulation-setup}
%%%%%%%%%%%%%%%%%%%%%%%%%%%%%%%%%%%%%

% \vspace{0.1cm}
\noindent{\em Physical network:} 
Three network topologies are considered. Two of them, \abilene\ and  \costnet, are taken from the SNDlib library \cite{orlowski2010sndlib} and the third is the fat-tree topology. The algorithms are evaluated in two network scales: ($i$) small-scale networks with the \abilene\ and a $2$-ary fat-tree topology and ($ii$) large-scale networks with the \costnet\ and a $6$-ary fat-tree topology.

\begin{figure}[b]
    \centering
    \includegraphics[width=0.8\columnwidth]{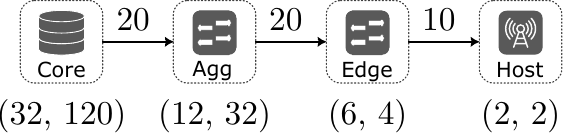}
    \caption{Four layers of the fat-tree topology: Each node provides a fixed amount of compute and storage resources, measured in the number of vCPUs and GBs respectively. Links have a fixed amount of available bandwidth, measured in Gbps.}
    \label{fig:fat-tree}
\end{figure}

A typical fat-tree has three layers: the core, aggregation, edge and host, which correspond to the data center, regional cloud, edge cloud and base stations (or radio units in 5G), as illustrated in Fig.~\ref{fig:fat-tree}.

\begin{figure}[tb]
    \centering
    \subfloat[Fat-tree. \label{fig:twofat-net}]{              
        \includegraphics[width=0.32\columnwidth]{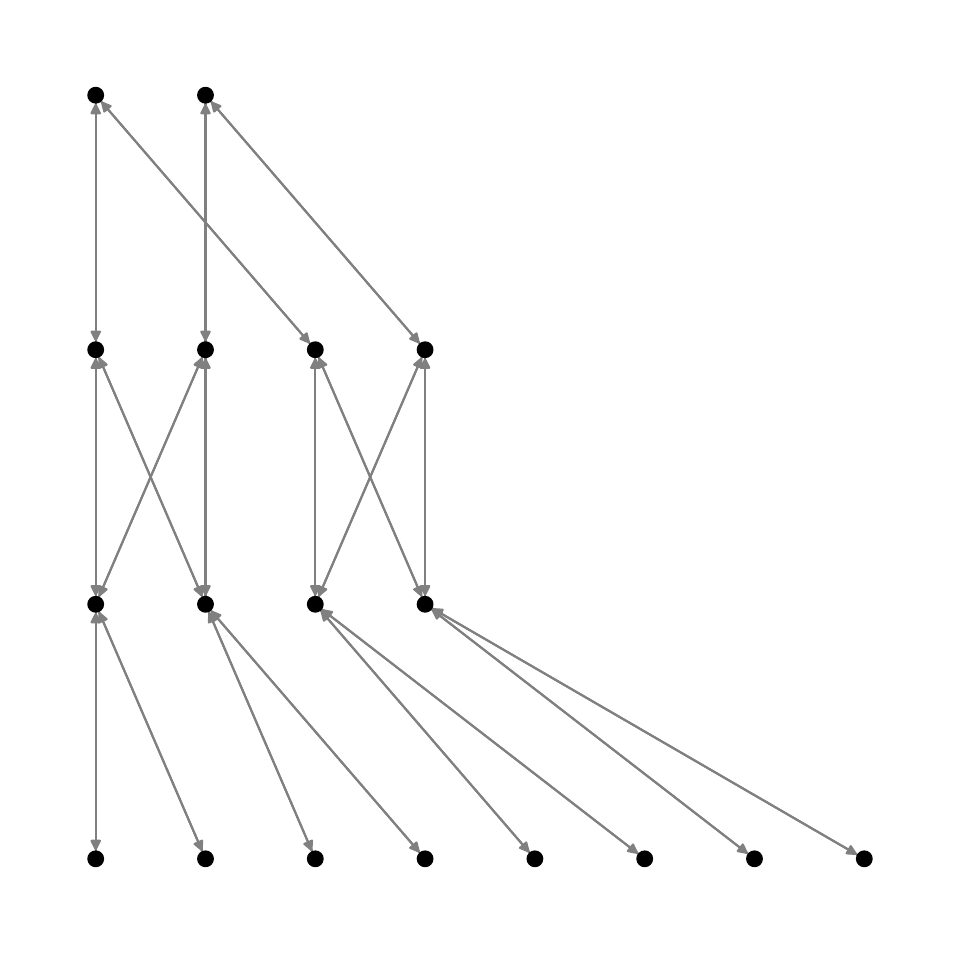}
    }
    \subfloat[\abilene. \label{fig:abilene-net}]{              
        \includegraphics[width=0.32\columnwidth]{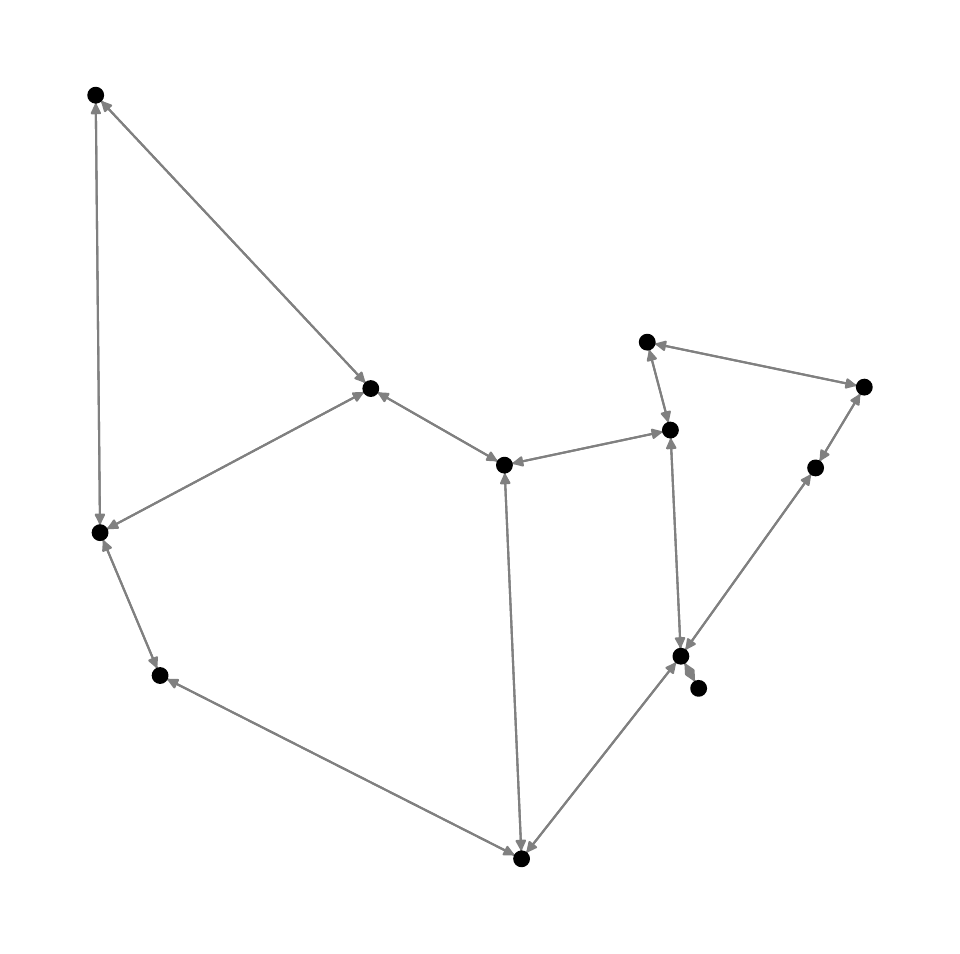}
    }
    \subfloat[\costnet. \label{fig:cost266-net}]{
        \includegraphics[width=0.32\columnwidth]{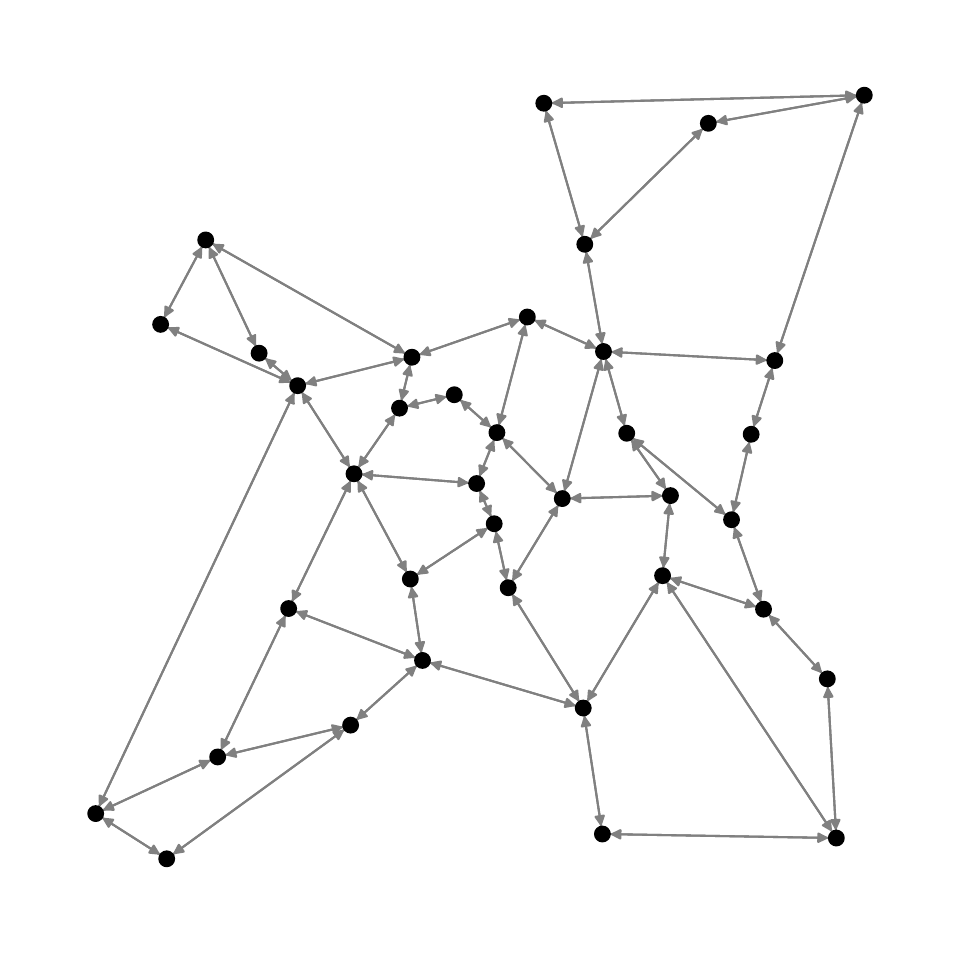}
    }
    \caption{Topology of the networks used for simulations: (a) A typical fat-tree topology ($2$-ary), (b) the \abilene, and (c) the \costnet\ topologies from the SNDlib library \cite{orlowski2010sndlib}.}
    \label{fig:abilene-cost266}
\end{figure}

A $2$-ary fat-tree consists of $2$ core nodes, $4$ aggregation nodes, $4$ edge nodes and $8$ host nodes (as illustrated in Fig.~\ref{fig:twofat-net}). For a $6$-ary network, the number of nodes at each layer is $9$, $18$, $18$, and $54$. 
In both fat-tree networks, each host, edge, aggregate, and core node has $2$, $6$, $12$, $32$ vCPUs and $2$, $4$, $32$, $120$ GBs of storage respectively. The bandwidth of each host-edge, edge-aggregate, and aggregate-core link are $10$ Gbps, $20$ Gbps, and $20$ Gbps, respectively.

\begin{table}[t]
    \caption{Characteristics of the Used Network Topologies}
    \label{tab:Networks}
    \centering
    % \vspace{-0.2cm}
    \begin{tabular}{l|cc}
        \toprule 
        \textbf{Network} & $|\cN|$ & $|\cL|$ \\
        \cmidrule[0.4pt](lr{0.12em}){1-1}%
        \cmidrule[0.4pt](lr{0.12em}){2-2}%
        \cmidrule[0.4pt](lr{0.12em}){3-3}%
        \abilene & $12$  & $30$    \\
        \costnet & $37$  & $114$   \\
        $2$-ary fat-tree & $18$ & $40$   \\
        $6$-ary fat-tree & $99$ & $324$  \\ 
        \bottomrule
    \end{tabular}
\end{table}

The topologies of the \abilene\ and  \costnet\ networks are depicted in Figs.~\ref{fig:abilene-net} and \ref{fig:cost266-net}. In both networks, each node provides $8$ vCPUs and $64$ GB of storage, whereas each link provides a bandwidth of $25$ Gbps.  Table~\ref{tab:Networks} summarizes the size of each physical network used for our simulations.
 
\vspace{0.1cm}
\noindent{\em Simulation scenarios:} We considers two following scenarios:
\begin{itemize}
    \item \textbf{Small-scale scenario}: Using \twofat\ and \abilene\ network topologies and $\vert\mathcal{S}\vert=15$ slices. The performance of \ilpse, \bnbstarinf, and \bfnse is compared;
    \item \textbf{Large-scale scenario}: Using \sixfat\ and \costnet\ network topologies and $\vert\mathcal{S}\vert=75$ slices. The performance of \bnbstarinf, \bnbstarlim{3}, and \bfnse is compared. The \ilpse\ is not considered in this scenario due to its complexity.

\end{itemize}

\begin{figure}[htb]
    \centering
    \subfloat[Configuration 1 ($k_1$). \label{fig:config-k1}]{
        \includegraphics[width=0.8\columnwidth]{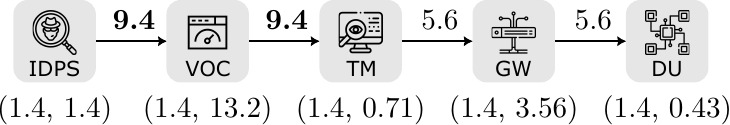}
    }\\
    \subfloat[Configuration 2 ($k_2$). \label{fig:config-k2}]{
        \includegraphics[width=0.8\columnwidth]{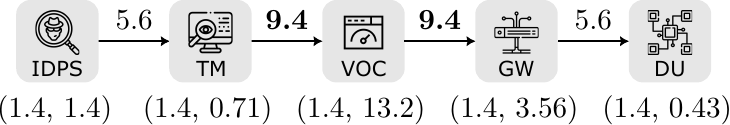}
    }
    \caption{The two network slice configurations used in the simulation. Each node requires a specific amount of compute and storage resources, measured in terms of vCPUs and GBs, respectively. Additionally, virtual links between VNFs have bandwidth requirements, measured in Gbps.}
    \label{fig:slices-setup}
\end{figure}

\vspace{0.1cm}
\noindent{\em Slices:} 
In all simulations, we consider the example slice with two possible configurations shown in Fig.~\ref{fig:flexible-example}. Each instance of this slice can deliver high-definition video streaming service to multiple user equipments. The resource demands of each configuration are detailed in Fig.~\ref{fig:slices-setup}. 

In both small-scale and large-scale scenarios, the benefit of allowing VNF order flexibility is investigated  by comparing the performance of the algorithms (\ilpse, \bnbstar\ and \bfnse) in three slice settings: 
\begin{itemize}
    \item \textbf{$\boldsymbol{k_1}$-only}: The configuration of all slices is fixed to $k_1$;
    
    \item \textbf{$\boldsymbol{k_2}$-only}: The configuration of all slices is fixed to $k_2$; 
    
    \item \textbf{flexible}: The VNO can flexibly select either $k_1$ or $k_2$ to deploy the slices. 
\end{itemize}

\vspace{0.1cm}
\noindent{\em Software and hardware:}
All simulations are performed on a PC equipped with an Intel i5-12500T CPU and $32$ GB of RAM. All algorithms are implemented using Python 3.10.14. \ilpse\ uses Gurobi Optimizer v11.0.3 as the ILP solver.

\vspace{0.1cm}
\noindent{\em Metrics:}
Two metrics are used to perform the evaluation: 
($i$) the acceptance rate (\emph{i.e.}, ${N(\boldsymbol{\pi})}/{\vert\mathcal{S}\vert}$)
and ($ii$) the total time used to obtain the results.

\vspace{0.1cm}
\noindent{\em Other parameters:}
For \ilpse, the tuning parameter $\gamma$ in the objective function of problem~\eqref{prob:SEflexorder} is set to $0.999$, as the value of $H(\boldsymbol{x})$ (number of used physical links) is typically much larger than that of $N(\boldsymbol{\pi})$ (number of accepted slices). For the A* cost function of \bnbstar, equal contributions of node and link usage (\textit{i.e.}, $\rho_{1}=\rho_{2}$) are considered in \eqref{eq:g-func} and \eqref{eq:h-func}.

%%%%%%%%%%%%%%%%%%%%%%%%%%%%%%%%%%%%%
\subsection{Results on Small-Scale Scenarios}
\label{subsec:smallscale}
%%%%%%%%%%%%%%%%%%%%%%%%%%%%%%%%%%%%%

%%%%%%%%%%%%%%%%%%%%%%%%%%%%%%%%%%%%%
% SUMMARY OF RESULTS (DO NOT DELETE IT)
% https://husteduvn-my.sharepoint.com/:x:/g/personal/trung_luuquang_hust_edu_vn/EXVRlVBYcr5HnzXj4u4wGAQBDg2sNm-NnmznLQ1vu22UeA?e=aJuwwb

% Results on Small-Scale Scenarios				
% 2-ary fat-tree						
% Config	ILP	BnB	BFN	ILP	BnB	BFN
% k1	11	7	6	73.3%	46.7%	40.0%
% k2	11	6	4	73.3%	40.0%	26.7%
% flex	12	7	7	80.0%	46.7%	46.7%
% Difference		6.7%	6.7%	20.0%
						
% abilene						
% Config	ILP	BnB	BFN	ILP	BnB	BFN
% k1	11	10	9	73.3%	66.7%	60.0%
% k2	11	11	9	73.3%	73.3%	60.0%
% flex	12	11	11	80.0%	73.3%	73.3%
% Difference		6.7%	6.7%	13.3%
						
% Results on Large-Scale Scenarios			
% 6-ary fat-tree						
% Config	ILP	BnB	BFN	ILP	BnB	BFN
% k1	51	30	39	68.0%	40.0%	52.0%
% k2	55	40	31	73.3%	53.3%	41.3%
% flex	59	50	31	78.7%	66.7%	41.3%
% Difference		10.7%	26.7%	0.0%
						
% abilene						
% Config	ILP	BnB	BFN	ILP	BnB	BFN
% k1	36	14	33	48.0%	18.7%	44.0%
% k2	34	34	22	45.3%	45.3%	29.3%
% flex	36	36	32	48.0%	48.0%	42.7%
% Difference		2.7%	29.3%	13.3%
%%%%%%%%%%%%%%%%%%%%%%%%%%%%%%%%%%%%%

% \vspace{0.1cm}

% ACCEPTANCE RATE -- small
\begin{figure}[htb]
    \centering
    \subfloat[\twofat\label{fig:2fat-case-compare-acceptance}]{
    \includegraphics[width=0.23\textwidth]{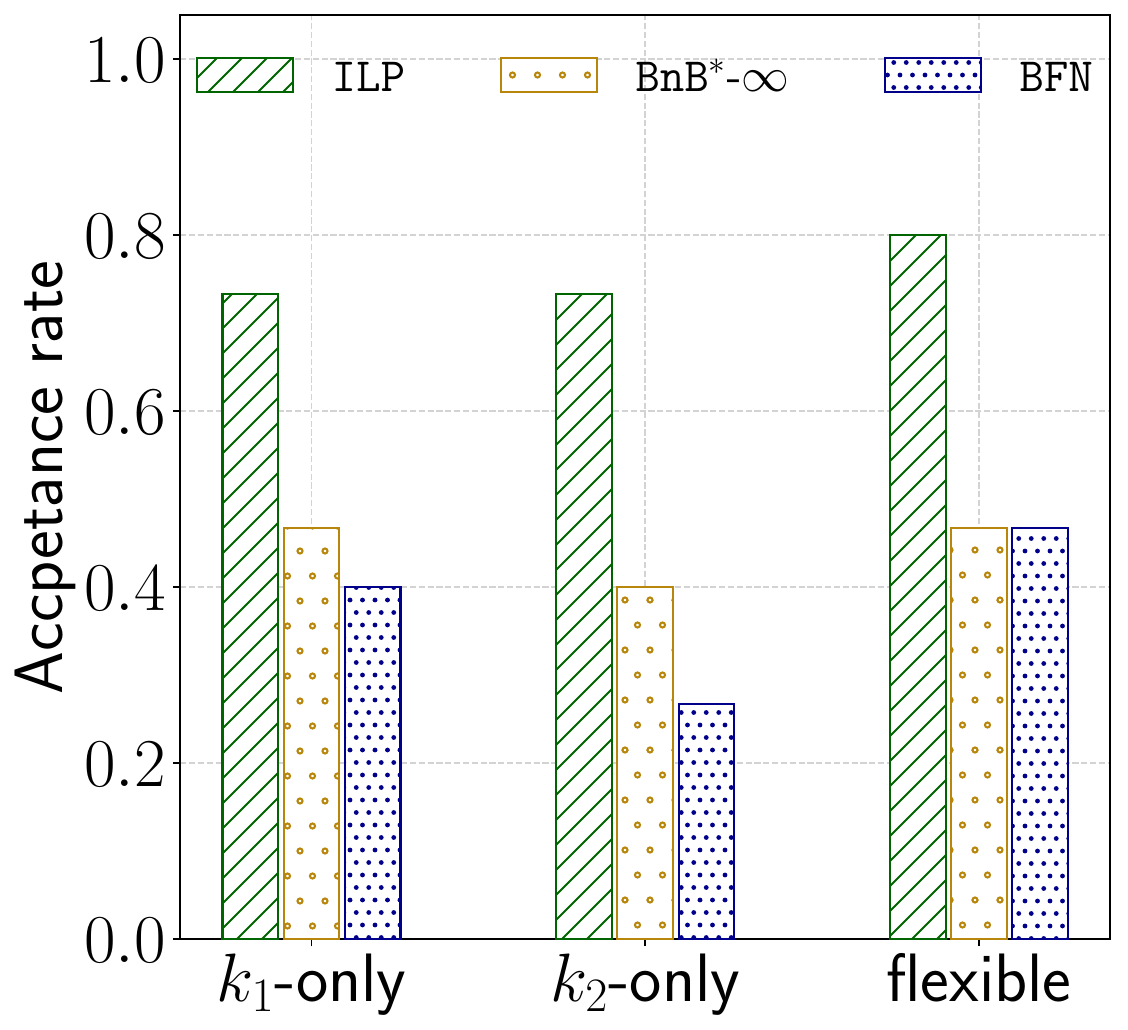}
    }
    \subfloat[\abilene\label{fig:abi-case-compare-acceptance}]{
    \includegraphics[width=0.23\textwidth]{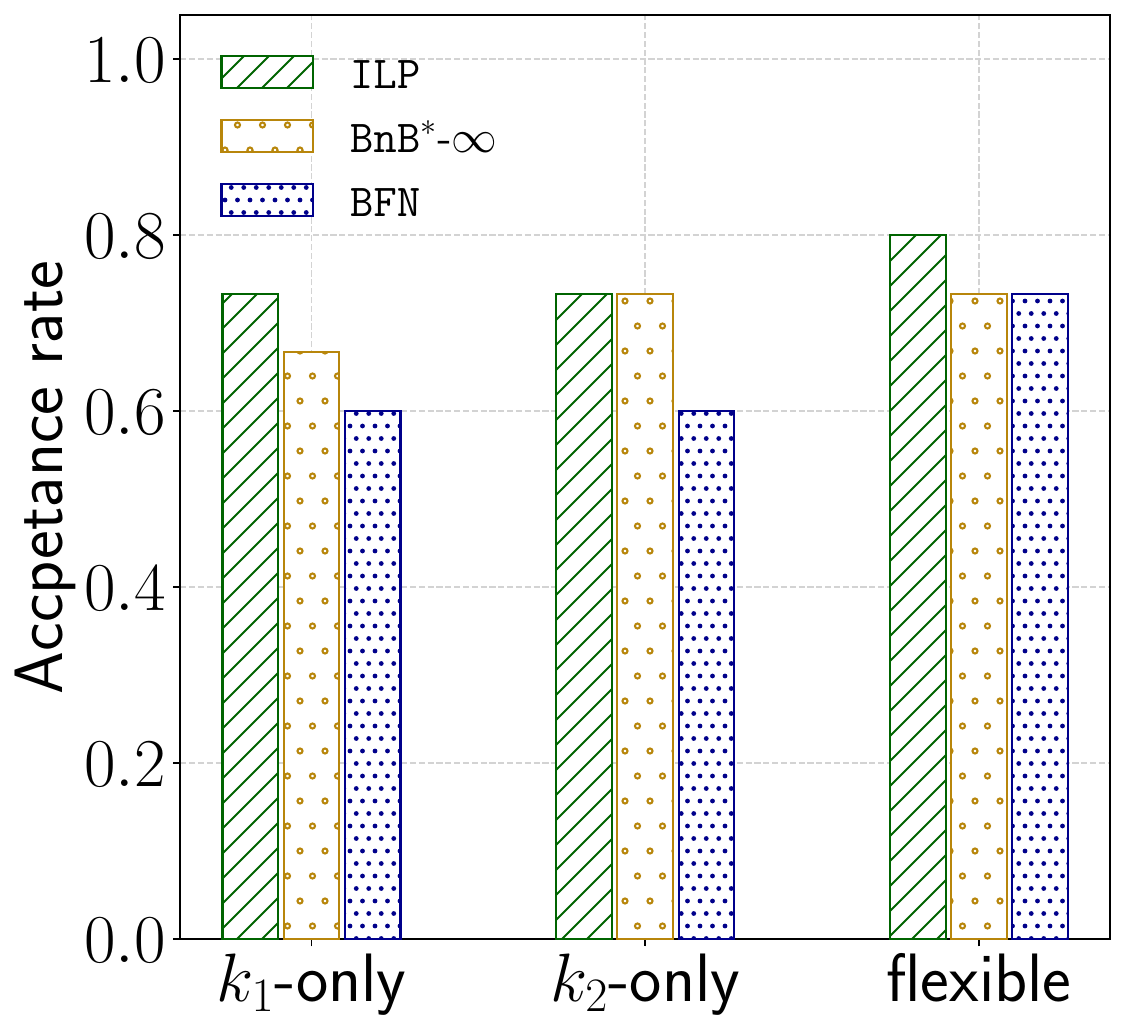}
    }
    \caption{Acceptance rate results in small-scale scenarios when using (a) \twofat\ and (b) \abilene\ network topology.}% \edit{REMOVE \bnbstarlim{3} FROM THIS FIGURE!}}
    \label{fig:small-acceptance-rate}
\end{figure}

\noindent{\bf Slice acceptance rate:}
Fig.~\ref{fig:small-acceptance-rate} depicts the slice acceptance rate for two network topologies: a $2$-ary fat-tree and the \abilene\ network, under different methods, including \ilpse, \bnbstarinf, and \bfnse. 

In all scenarios and with all methods, VNF order flexibility brings improvement in terms of acceptance rate, with a performance gap of nearly $7\%$ for \ilpse\ and \bnbstarinf, compared to that with the $k_1$-only and $k_2$-only settings. \bfnse\ even achieves a higher performance gap, $13.3\%$ with the $2$-ary fat-tree network and $20\%$ with the \abilene\ network.

Moreover, our proposed algorithm (\bnbstarinf) showcases a relatively good performance compared to that of \ilpse, especially in the results with the \abilene\ network, where \bnbstarinf\ yields an acceptance rate close to that of \ilpse\ ($73.3\%$ compared to $80\%$). For instance, in the $k_1$-only setting and using the \abilene\ network, the gap between \bnbstarinf\ and \bfnse\ is $6.7\%$, whereas in the $k_2$-only setting, this gap becomes $13.3\%$.

\bfnse\ also demonstrates the capability of leveraging VNF order flexibility to increase the acceptance rate. It obtains a similar performance to that of \bnbstarinf\ in the flexible setting and in both network topologies. 

Finally, as anticipated, \ilpse\ consistently 
achieves the highest acceptance rates across all configurations and settings.
\ilpse\ achieves an acceptance rate of around $70\%$--$80\%$ across all three settings and network topologies---demonstrating its efficiency, especially when dealing with the $2$-ary fat-tree topology, wherein both \bnbstarinf\ and \bfnse\ achieve a much lower acceptance rate of around $47\%$.

% This result aligns with expectations, as \ilpse's optimal approach is particularly effective in small-scale scenarios wherein exact solutions are computationally feasible.

% In the $2$-ary fat-tree topology, \ilpse\ achieves an acceptance rate of around \hidden{$80\%$} \edit{$70\% - 80\%$} across all three configurations-$k_1$-only, $k_2$-only, and flexible-demonstrating its efficiency in handling network slicing requests within this structured, tree-based topology. \bnbstarinf\ also performs relatively well, outperforming \bfnse\ overall, particularly in the $k_1$-only and $k_2$-only settings, with performance gains of \hidden{$5\%$ and $15\%$} \edit{$6.7\%$ and $13.3\%$}, respectively. In the flexible configuration, all algorithms show improved slice acceptance rates compared to the $k_1$-only and $k_2$-only settings, where each slice configuration is unique. Specifically, \ilpse\ achieves a \hidden{$5\%$} \edit{$6.7\%$} performance improvement when embedding slices with flexible VNF ordering. This highlights the advantage of permitting VNF order flexibility, which results in a higher slice acceptance rate.

% A similar performance trend is observed in simulations with the \abilene\ topology. In this setup, \ilpse\ maintains a comparable acceptance rate to that achieved in the 2-ary fat-tree network. Notably, \bnbstarinf\ performs very well, narrowing its performance gap with \ilpse\ to just \hidden{$5\%$} \edit{$6.7\%$}, as opposed to the \hidden{$15\%$} \edit{$33.3\%$} gap observed in the 2-ary fat-tree network.

% RUNTIME -- small
\begin{figure}[htb]
    \centering
    \subfloat[\twofat. \label{fig:2fat-case-compare-runtime}]{
    \includegraphics[width=0.23\textwidth]{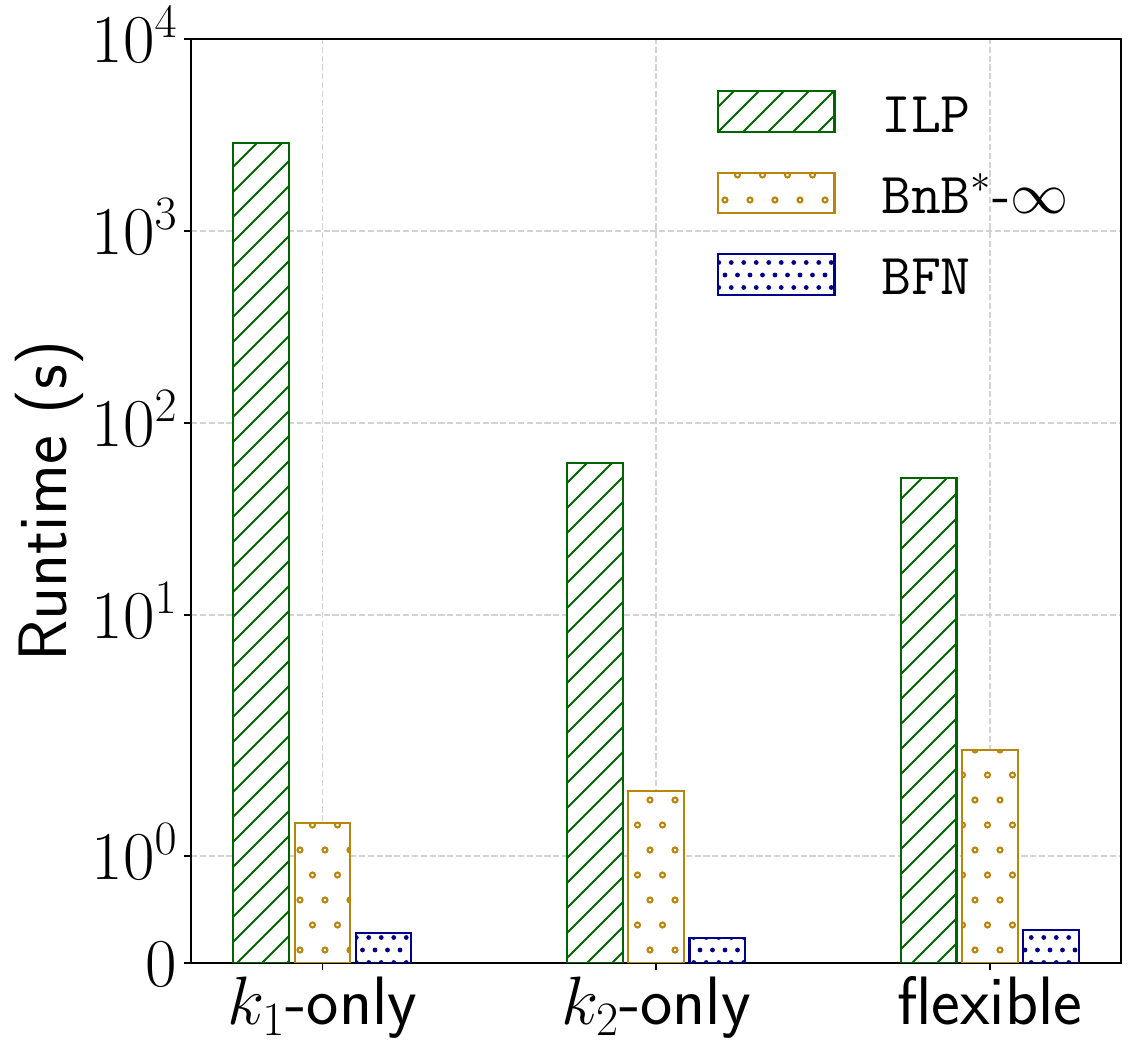}
    }
    \subfloat[\abilene. \label{fig:abi-case-compare-runtime}]{
    \includegraphics[width=0.23\textwidth]{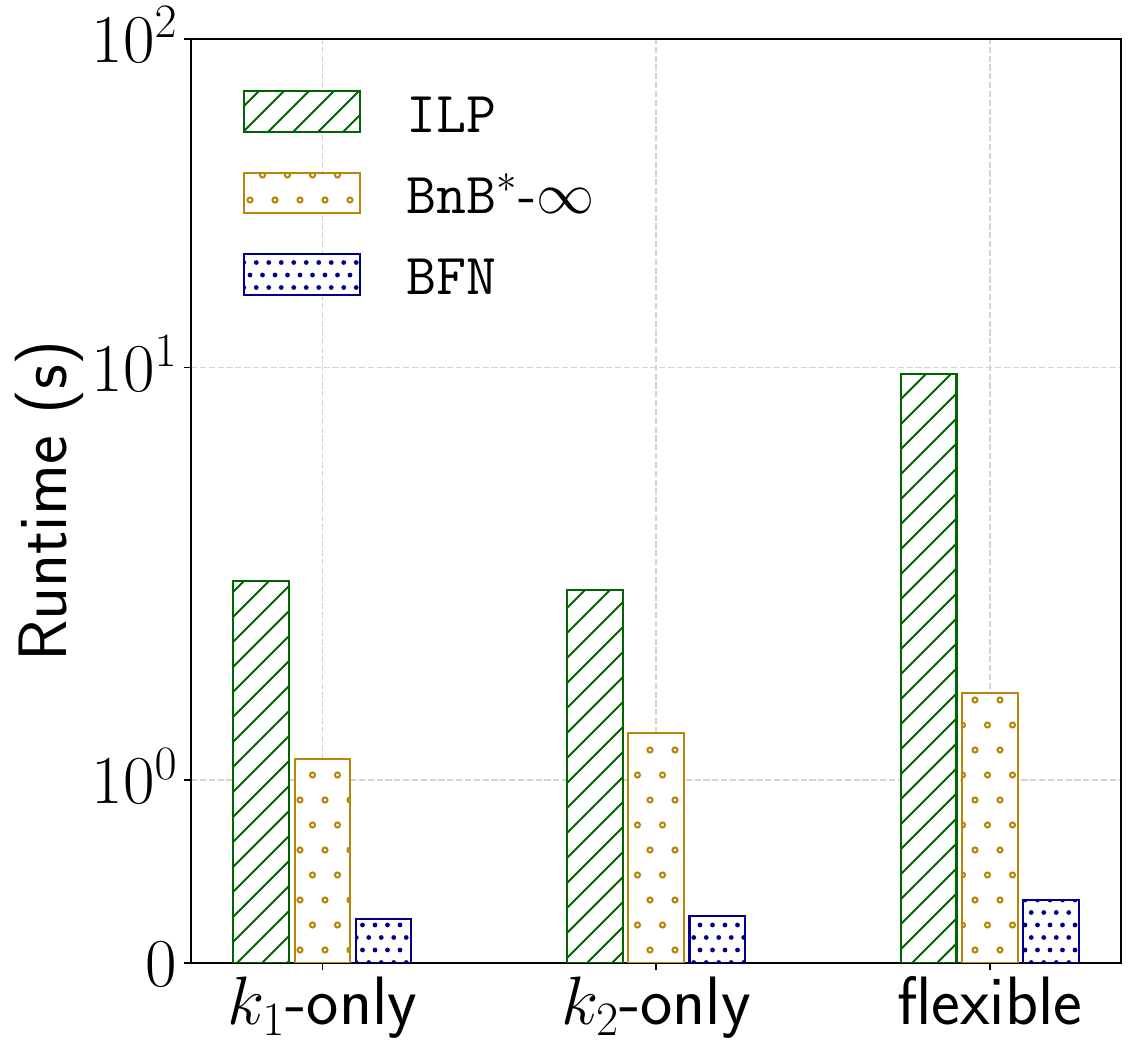}
    }
    \caption{Runtime (in seconds) results in small-scale settings when using (a) \twofat\ and (b) \abilene\ network topology.} % \edit{REMOVE \bnbstarlim{3} FROM THIS FIGURE!}}
    \label{fig:small-runtime}
\end{figure}

\vspace{0.1cm}
\noindent{\bf Runtime:}
Fig.~\ref{fig:small-runtime} presents the runtime results of the algorithms. As expected, \ilpse\ exhibits the longest runtime, orders of magnitude higher than the other methods across all scenarios and configurations. This is due to its exact optimization approach, which becomes computationally expensive even in small-scale settings.
\bfnse\ is the fastest method, requiring only a few seconds to complete slice embeddings across all scenarios. \bnbstarinf\ demonstrates moderate runtimes. The runtime increase in the flexible setting is minimal compared to the $k_1$-only and $k_2$-only configurations.

Overall, these results highlight a clear trade-off between runtime and solution quality. \ilpse\ provides the best acceptance rates but at the cost of significantly longer runtimes. \bnbstarinf\ offers a reasonable compromise, providing better runtimes than \ilpse\ while still maintaining competitive acceptance rates. \bfnse, though computationally efficient, sacrifices solution quality.
Most importantly, the results in this small-scale scenario confirm the advantage of allowing VN ordering flexibility in improving the slice acceptance rate.

%%%%%%%%%%%%%%%%%%%%%%%%%%%%%%%%%%%%%
\subsection{Results on Large-Scale Scenarios}
\label{subsec:largescale}
%%%%%%%%%%%%%%%%%%%%%%%%%%%%%%%%%%%%%

%In this large-scale scenario, we evaluate the performance of two variants of \bnbstar\ (\bnbstarinf\ and \bnbstarlim{3}) and compare them with that of \bfnse. This evaluation provides insight into the capability of \bnbstar\ for solving larger instances of Problem~\eqref{prob:SEflexorder}.

\begin{figure}[htb]
    \centering
    % ACCEPTANCE -- large
    \subfloat[\sixfat. \label{fig:6fat-case-compare-acceptance}]{
    \includegraphics[width=0.23\textwidth]{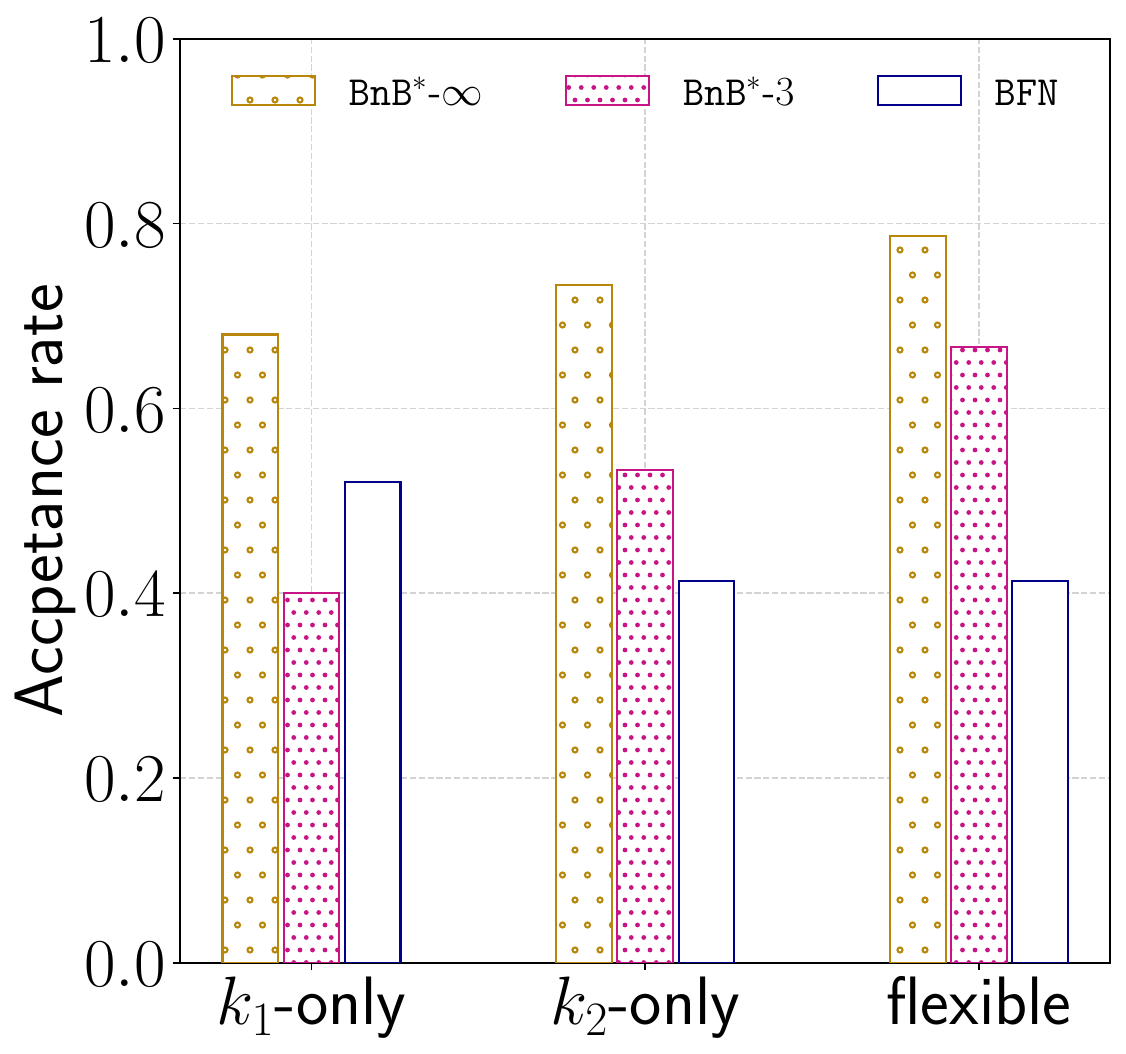}
    }
    \subfloat[\costnet. \label{fig:cost266-case-compare-acceptance}]{
    \includegraphics[width=0.23\textwidth]{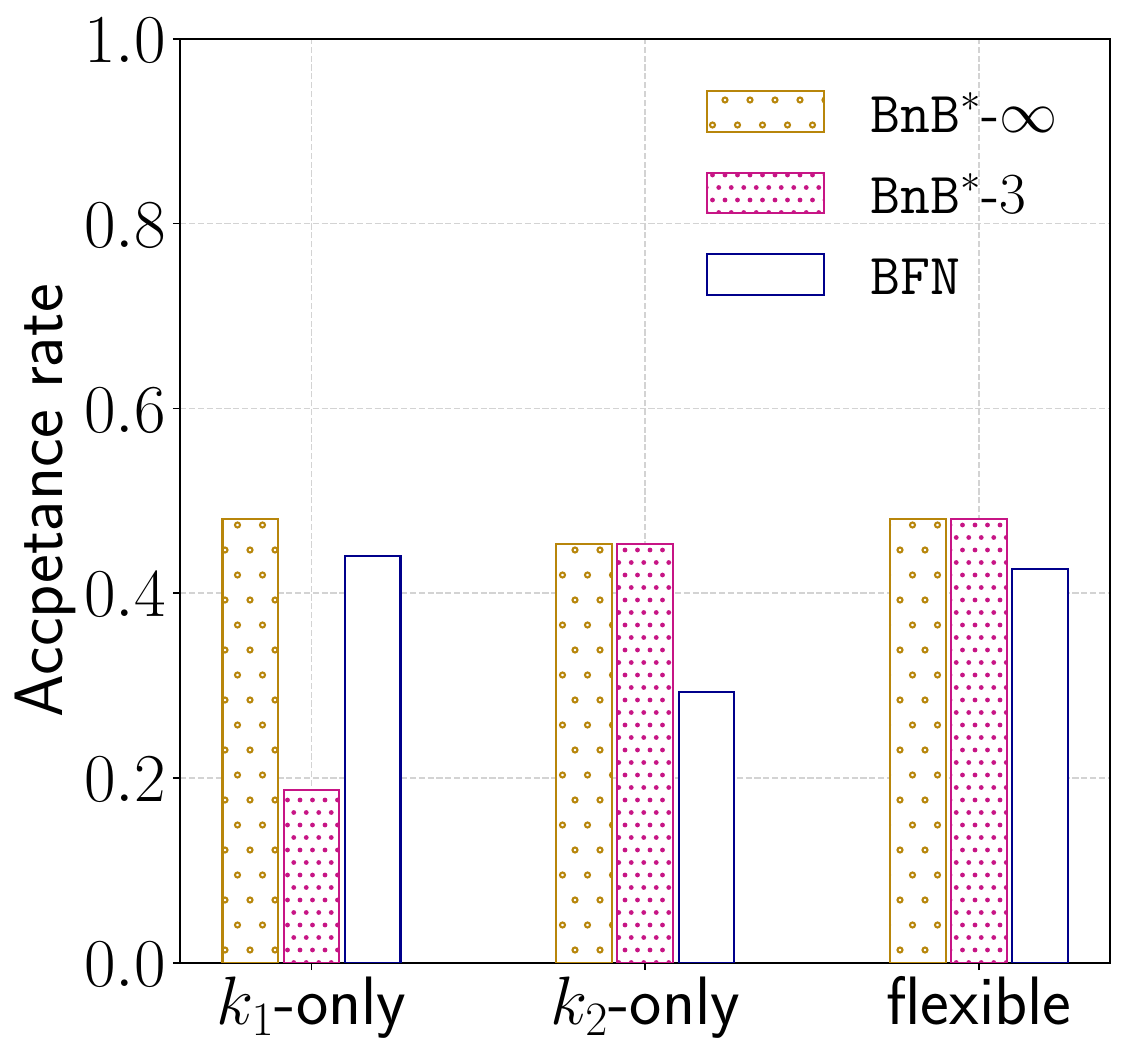}
    }
    \caption{Acceptance rate in large-scale settings  when using (a) \twofat\ and (b) \abilene\ network topology.}
    \label{fig:large-acceptance-rate}
\end{figure}

\vspace{0.1cm}
\noindent{\bf Slice acceptance rate:}
Recall that, in the small-scale scenarios, \bfnse\ performs relatively well, achieving acceptance rates comparable to those of \bnbstarinf. Notably, in the flexible setting, \bfnse\ can even match the acceptance rate of \bnbstarinf\ (see again Section~\ref{subsec:smallscale}).
In the large-scale scenarios using the $6$-ary fat-tree and \costnet\ topologies, a clear shift in performance patterns emerges, as shown in Fig.~\ref{fig:large-acceptance-rate}. \bnbstarinf\ consistently achieves the highest slice acceptance rate among the three algorithms, with a substantial performance gap over \bfnse\ in the flexible setting, amounting to around $37.3\%$. The performance of \bfnse\ notably declines in these large-scale scenarios compared to the small-scale results, particularly in the $k_2$-only setting. Additionally, \bfnse\ does not show any improvement in slice acceptance rate when run in the flexible setting, indicating its limited effectiveness when compared to the BnB-based approaches.

In contrast, the limited-complexity variant, \bnbstarlim{3}, demonstrates performance comparable to its unlimited counterpart, \bnbstarinf, especially in the flexible setting. In the 6-ary fat-tree topology, \bnbstarlim{3} achieves an acceptance rate of $66.7\%$, close to that achieved by \bnbstarinf\ ($78.7\%$). Notably, in the \costnet\ network, \bnbstarlim{3} reaches the same acceptance rate as \bnbstarinf\ ($48\%$), outperforming \bfnse, which achieves a rate of $42.7\%$.

\begin{figure}[htb]
    \centering
    % RUNTIME -- large
    \subfloat[\sixfat. \label{fig:6fat-case-compare-runtime}]{
    \includegraphics[width=0.23\textwidth]{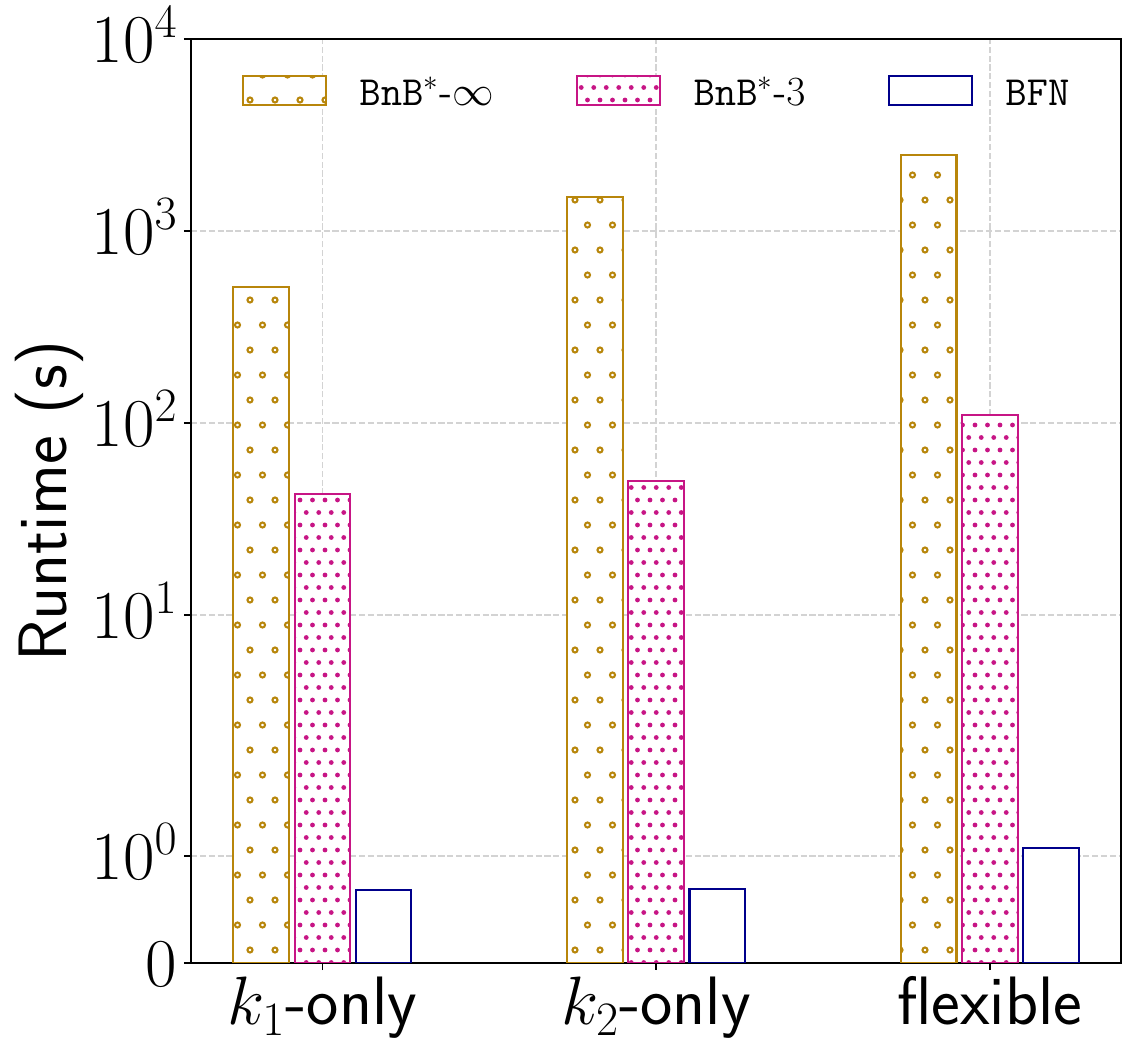}
    }
    \subfloat[\costnet. \label{fig:cost266-case-compare-runtime}]{
    \includegraphics[width=0.23\textwidth]{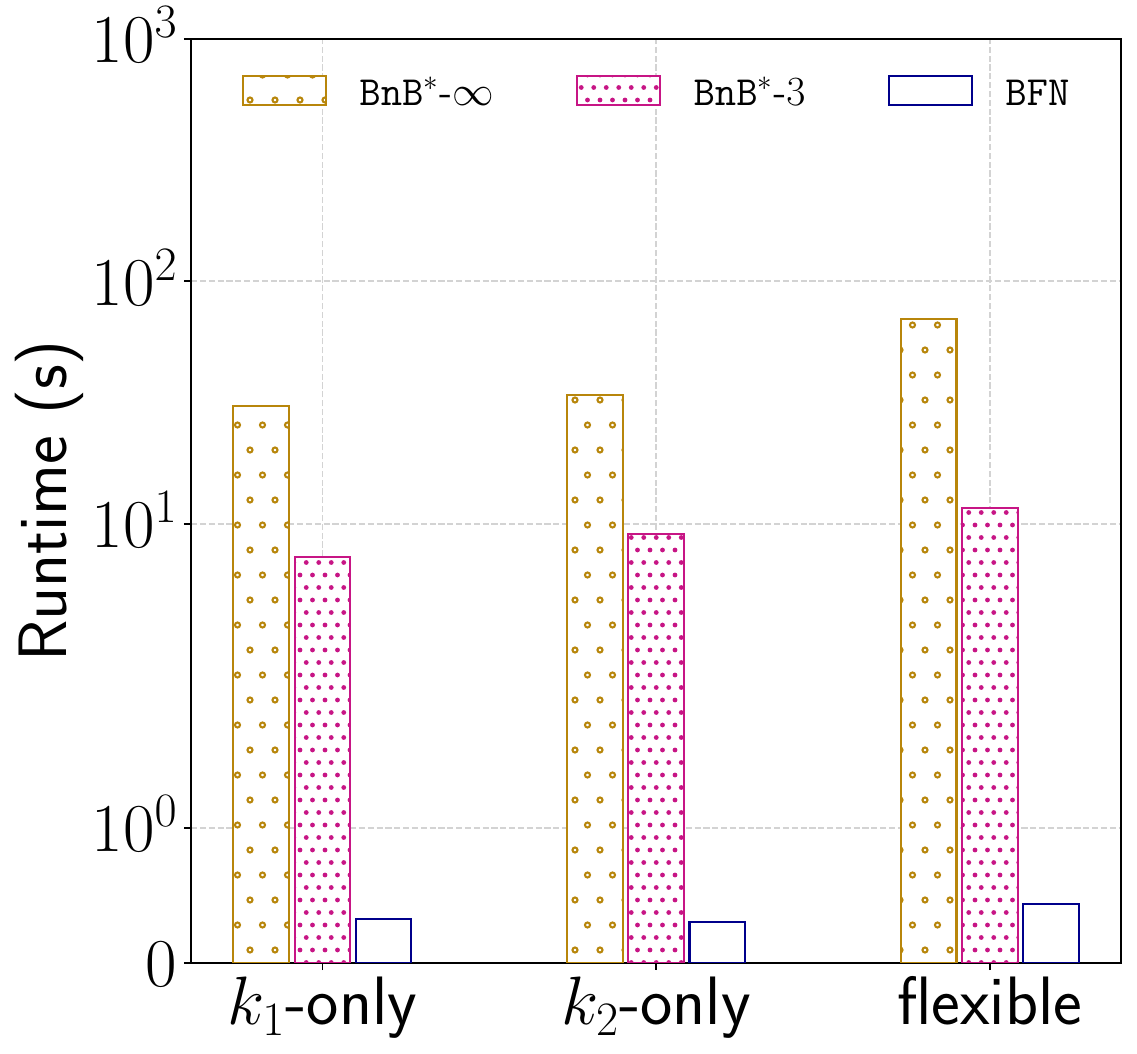}
    }
    \caption{Runtime (in seconds) in large-scale settings when using (a) \twofat\ and (b) \abilene\ network topology.}
    \label{fig:large-runtime}
\end{figure}

\vspace{0.1cm}
\noindent{\bf Runtime:}
Fig.~\ref{fig:large-runtime} shows the runtime performance in large-scale scenarios, which naturally increases compared to the small-scale scenarios (Fig.~\ref{fig:small-runtime}). \bnbstarinf\ remains the most computationally expensive method, with runtime reaching $10^2$ seconds, especially in the flexible setting for both the $6$-ary fat-tree and \costnet\ topologies. \bnbstarlim{3} and BFN continue to demonstrate faster runtimes, with BFN being the most efficient across all configurations. Interestingly, \bnbstarinf\ sees only a modest runtime increase in the flexible setting compared to the $k_1$-only and $k_2$-only settings. \bnbstarlim{3}, meanwhile, strikes a reasonable balance between acceptance rate and runtime. In contrast, although BFN achieves the fastest computation times, its acceptance rates consistently suffer, particularly as network complexity rises.

%%%%%%%%%%%%%%%%%%%%%%%%%%%%%%%%%%%%%
\subsection{Effect of Configuration Selection}
%%%%%%%%%%%%%%%%%%%%%%%%%%%%%%%%%%%%%

\begin{figure}[htb]
    \centering
    % ACCEPTANCE -- FLEX
    \subfloat[Small-scale scenarios.\label{fig:small-flex-acceptance-rate}]{
    \includegraphics[width=0.23\textwidth]{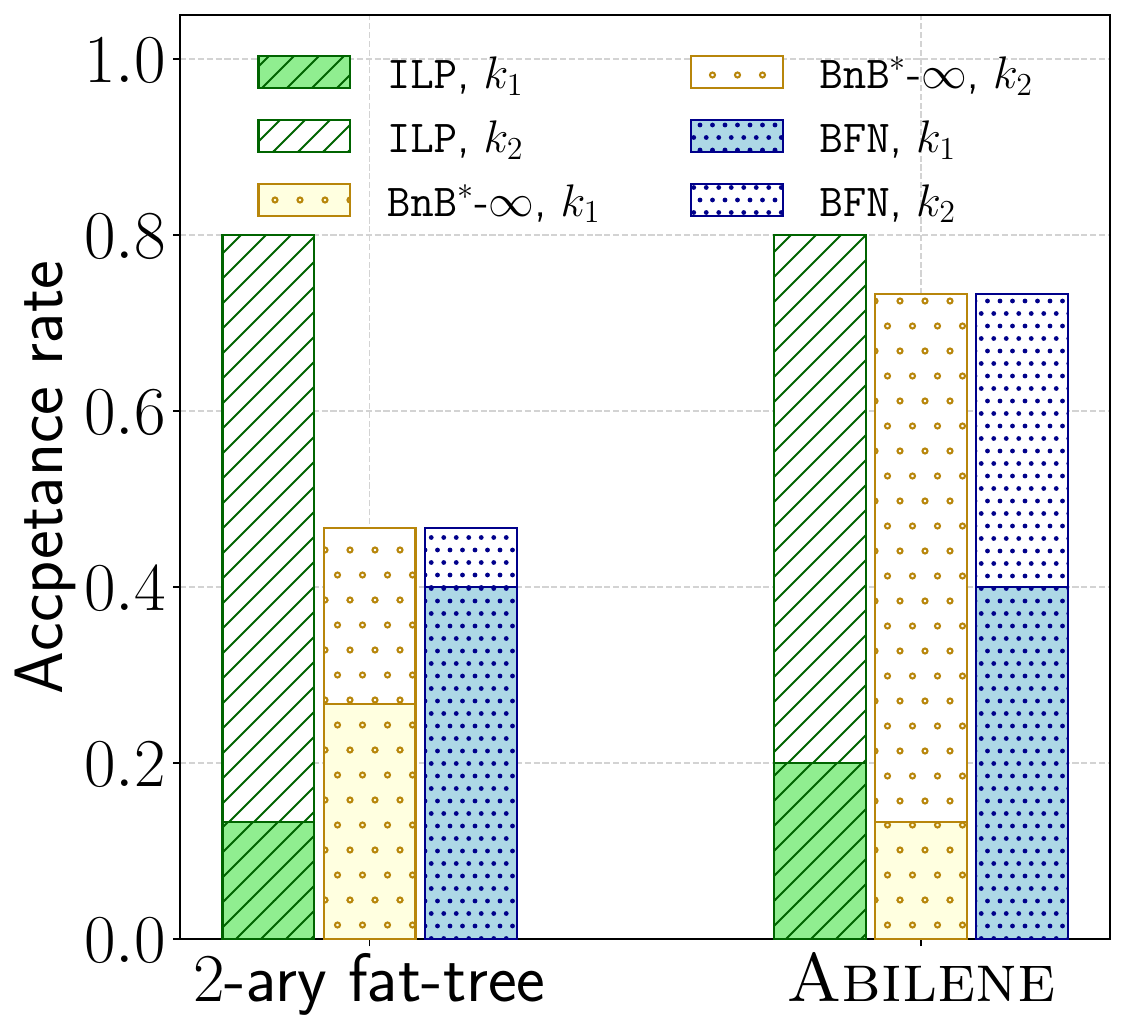}
    }
    \subfloat[Large-scale scenarios.\label{fig:large-flex-acceptance-rate}]{
    \includegraphics[width=0.23\textwidth]{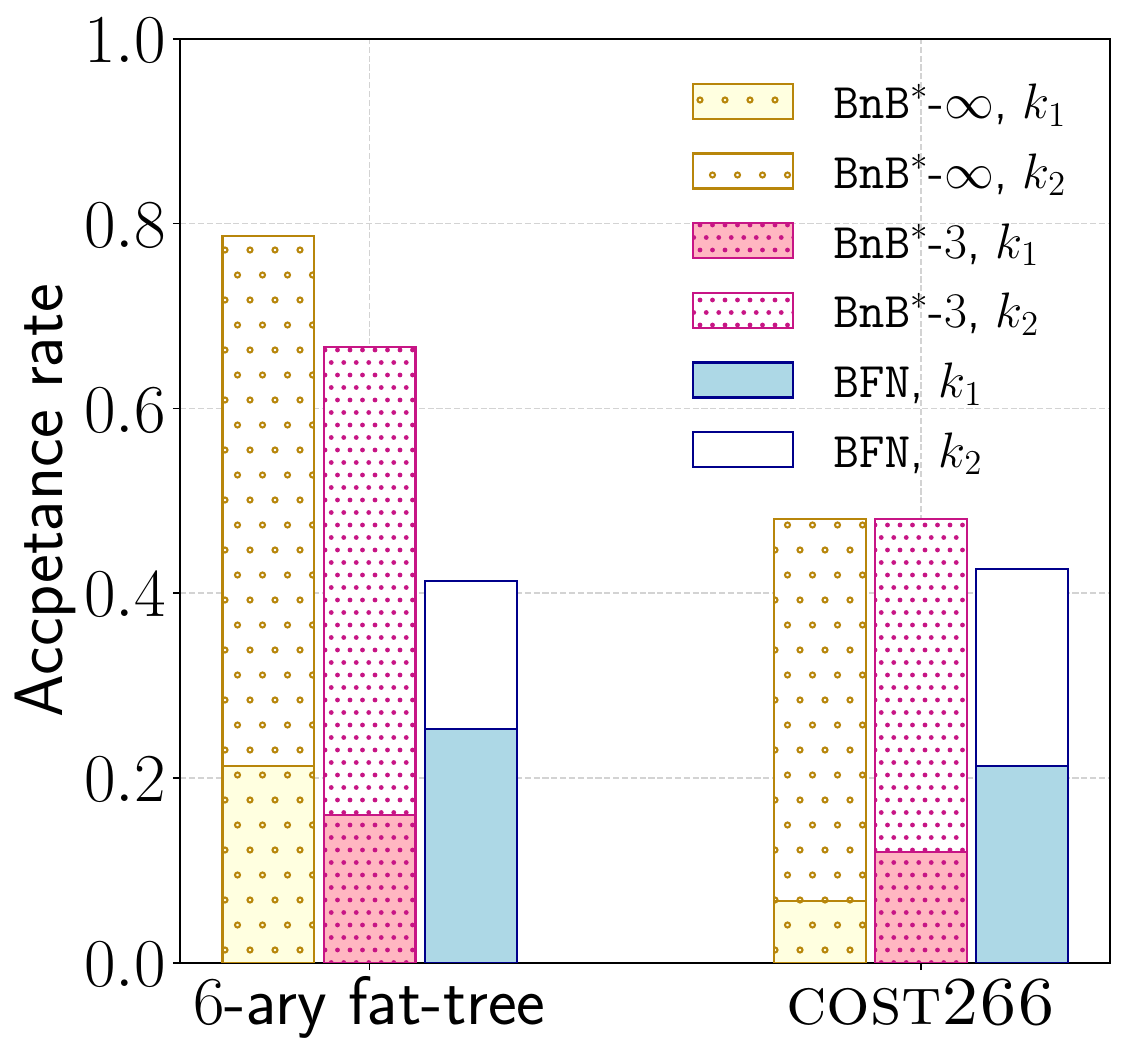}
    }
    \caption{Acceptance rate per configuration using flexible settings in (a) small-scale and (b) large-scale scenarios.} % \edit{Modify $X$-fattree to $X$-ary fat-tree}}
    \label{fig:flex-acceptance-rate}
\end{figure}

The results presented in Fig.~\ref{fig:flex-acceptance-rate} analyze the contribution of each slice configuration (\textit{i.e.}, $k_1$ and $k_2$) to the total acceptance rates of various algorithms  across different network scales.

Fig.~\ref{fig:flex-acceptance-rate} indicates that, unlike \ilpse\ and \bnbstar, \bfnse\ tends to favor configuration $k_1$, as it has a cascade bandwidth requirements (see Fig.~\ref{fig:config-k1}), which is similar to the hierarchical bandwidth structure of the fat-tree topology.
% \footnote{This is known as \textit{oversubscription}, a widely adopted design in data center networks to mitigate
% network resource under-utilization and reduce operational cost \cite{al2008scalable}.}
In contrast, \ilpse, \bnbstarinf, and \bnbstarlim{3} achieve a better balance between the two configurations, resulting in a higher overall slice acceptance rate.

%%%%%%%%%%%%%%%%%%%%%%%%%%%%%%%%%%%%%
\section{Conclusion} 
\label{sec:Conclusion}
%%%%%%%%%%%%%%%%%%%%%%%%%%%%%%%%%%%%%
In this paper, we have introduced an effective optimization framework for slice admission control and embedding within a network infrastructures, emphasizing flexibility in VNF order within network slices. The problem is formulated as an ILP to jointly optimize the admission control, VNF order selection, and embedding of network slices. Due to the computational intensity of solving ILP for large-scale networks, we have also proposed scalable sub-optimal solutions, including a BnB-based method (\bnbstar) and a greedy algorithm (\bfnse), both designed for efficient deployment in large networks.
Our simulations across multiple network topologies and slice configurations indicate that allowing flexible VNF ordering significantly enhances slice acceptance rates and improves resource utilization. This adaptable approach effectively addresses the growing demands of diverse digital services, providing scalability and efficiency beyond traditional static VNF configurations.

For future work, we plan to extend our study to more complex, realistic scenarios, exploring various network slice configurations and diverse topologies to better address the challenges of next-generation networks. 
Additionally, we aim to explore advanced machine learning techniques by employing various deep reinforcement learning (DRL) architectures to address problem~\eqref{prob:SEflexorder}. 

%Potential approaches include Q-learning \cite{haque2022network,thanh2024accelerating}, deep Q-learning \cite{doanis2022scalable,liu2024reinforcement}, dueling double deep Q-learning \cite{kwantwi2023blockchain}, which can separate value and advantage functions to improve learning stability. 
%Another promising option is the deep deterministic policy gradient (DDPG) algorithm \cite{mei2021intelligent,mai2021transfer}, known for its effectiveness in continuous action spaces.

% DDPG:  \cite{mei2021intelligent,mai2021transfer}

\bibliographystyle{IEEEtran}

\bibliography{biblio_trung,bibliography}

\begin{thebibliography}{10}
\providecommand{\url}[1]{#1}
\csname url@samestyle\endcsname
\providecommand{\newblock}{\relax}
\providecommand{\bibinfo}[2]{#2}
\providecommand{\BIBentrySTDinterwordspacing}{\spaceskip=0pt\relax}
\providecommand{\BIBentryALTinterwordstretchfactor}{4}
\providecommand{\BIBentryALTinterwordspacing}{\spaceskip=\fontdimen2\font plus
\BIBentryALTinterwordstretchfactor\fontdimen3\font minus
  \fontdimen4\font\relax}
\providecommand{\BIBforeignlanguage}[2]{{%
\expandafter\ifx\csname l@#1\endcsname\relax
\typeout{** WARNING: IEEEtran.bst: No hyphenation pattern has been}%
\typeout{** loaded for the language `#1'. Using the pattern for}%
\typeout{** the default language instead.}%
\else
\language=\csname l@#1\endcsname
\fi
#2}}
\providecommand{\BIBdecl}{\relax}
\BIBdecl

\bibitem{foukas2017network}
X.~Foukas, G.~Patounas, A.~Elmokashfi, and M.~K. Marina, ``Network slicing in
  {5G}: Survey and challenges,'' \emph{IEEE Commun. Mag.}, vol.~55, no.~5, pp.
  94--100, 2017.

\bibitem{rafique2024survey}
W.~Rafique, J.~Barai, A.~O. Fapojuwo, and D.~Krishnamurthy, ``A survey on
  beyond {5G} network slicing for smart cities applications,'' \emph{IEEE
  Commun. Surv. Tutor.}, pp. 1--1, 2024.

\bibitem{ebrahimi2024resource}
S.~Ebrahimi, F.~Bouali, and O.~C.~L. Haas, ``Resource management from
  single-domain 5g to end-to-end 6g network slicing: A survey,'' \emph{IEEE
  Commun. Surv. Tut.}, pp. 1--1, 2024.

\bibitem{su2019resource}
R.~Su, D.~Zhang, R.~Venkatesan, Z.~Gong, C.~Li, F.~Ding, F.~Jiang, and Z.~Zhu,
  ``Resource allocation for network slicing in {5G} telecommunication networks:
  A survey of principles and models,'' \emph{IEEE Netw.}, vol.~33, no.~6, pp.
  172--179, 2019.

\bibitem{dealwis2023survey}
C.~De~Alwis, P.~Porambage, K.~Dev, T.~R. Gadekallu, and M.~Liyanage, ``A survey
  on network slicing security: Attacks, challenges, solutions and research
  directions,'' \emph{IEEE Commun. Surv. Tut.}, vol.~26, no.~1, pp. 534--570,
  2024.

\bibitem{donatti2023survey}
A.~Donatti, S.~L. Corr\^ea, J.~S.~B. Martins, A.~J.~G. Abelem, C.~B. Both,
  F.~de~Oliveira~Silva, J.~A. Suruagy, R.~Pasquini, R.~Moreira, K.~V. Cardoso,
  and T.~C. Carvalho, ``Survey on machine learning-enabled network slicing:
  Covering the entire life cycle,'' \emph{IEEE Trans. Netw. Service Manag.},
  vol.~21, no.~1, pp. 994--1011, 2024.

\bibitem{saha2024survey}
N.~Saha, M.~Zangooei, M.~Golkarifard, and R.~Boutaba, ``Deep reinforcement
  learning approaches to network slice scaling and placement: A survey,''
  \emph{IEEE Commun. Mag.}, vol.~61, no.~2, pp. 82--87, 2023.

\bibitem{zangooei2024flexible}
M.~Zangooei, M.~Golkarifard, M.~Rouili, N.~Saha, and R.~Boutaba, ``Flexible ran
  slicing in open ran with constrained multi-agent reinforcement learning,''
  \emph{IEEE J. Sel. Areas Commun.}, vol.~42, no.~2, pp. 280--294, 2024.

\bibitem{Riggio2016}
R.~Riggio, A.~Bradai, D.~Harutyunyan, T.~Rasheed, and T.~Ahmed, ``Scheduling
  wireless virtual networks functions,'' \emph{IEEE Trans. Netw. Service
  Manag.}, vol.~13, no.~2, pp. 240--252, 2016.

\bibitem{bouten2017semantically}
N.~Bouten, R.~Mijumbi, J.~Serrat, J.~Famaey, S.~Latr{\'e}, and F.~De~Turck,
  ``Semantically enhanced mapping algorithm for affinity-constrained service
  function chain requests,'' \emph{IEEE Trans. Netw. Service Manag.}, vol.~14,
  no.~2, pp. 317--331, 2017.

\bibitem{Vizarreta2018qos}
P.~Vizarreta, M.~Condoluci, C.~M. Machuca, T.~Mahmoodi, and W.~Kellerer,
  ``{QoS}-driven function placement reducing expenditures in nfv deployments,''
  in \emph{Proc. IEEE International Conference on Communications (ICC)}, 2017,
  pp. 1--7.

\bibitem{luu2022admission}
Q.-T. Luu, S.~Kerboeuf, and M.~Kieffer, ``Admission control and resource
  reservation for prioritized slice requests with guaranteed {SLA} under
  uncertainties,'' \emph{IEEE Trans. Netw. Service Manag.}, vol.~19, no.~3, pp.
  3136--3153, 2022.

\bibitem{Zhang2023multi}
P.~Zhang, N.~Chen, S.~Li, K.-K.~R. Choo, C.~Jiang, and S.~Wu, ``Multi-domain
  virtual network embedding algorithm based on horizontal federated learning,''
  \emph{IEEE Trans. Inf. Forensics Secur.}, vol.~18, pp. 3363--3375, 2023.

\bibitem{thanh2024accelerating}
M.-T. Nguyen, Q.-T. Luu, T.-H. Nguyen, D.-M. Tran, T.-A. Do, K.-H. Do, and
  V.-H. Nguyen, ``Accelerating network slice embedding with reinforcement
  learning,'' in \emph{Proc. IEEE International Conference on Communications
  and Electronics (ICCE)}, 2024.

\bibitem{ocampo2017optimal}
A.~F. Ocampo, J.~Gil-Herrera, P.~H. Isolani, M.~C. Neves, J.~F. Botero,
  S.~Latr{\'e}, L.~Zambenedetti, M.~P. Barcellos, and L.~P. Gaspary, ``Optimal
  service function chain composition in network functions virtualization,'' in
  \emph{Proc. 11th IFIP WG 6.6 International Conference on Autonomous
  Infrastructure, Management, and Security (AIMS)}.\hskip 1em plus 0.5em minus
  0.4em\relax Springer International Publishing, 2017, pp. 62--76.

\bibitem{huang2015converged}
J.~Huang, Q.~Duan, S.~Guo, Y.~Yan, and S.~Yu, ``Converged network-cloud service
  composition with end-to-end performance guarantee,'' \emph{IEEE Trans. Cloud
  Comput.}, vol.~6, no.~2, pp. 545--557, 2015.

\bibitem{luu2020coverage}
Q.-T. Luu, S.~Kerboeuf, A.~Mouradian, and M.~Kieffer, ``A coverage-aware
  resource provisioning method for network slicing,'' \emph{IEEE/ACM Trans.
  Netw.}, vol.~28, no.~6, pp. 2393--2406, 2020.

\bibitem{luu2024flexorder}
Q.-T. Luu, M.-T. Nguyen, T.-H. Nguyen, M.~Kieffer, V.-D. Nguyen, Q.-L. Luu, and
  T.-T. Nguyen, ``Admission control and embedding of network slices with
  flexible {VNF} order,'' in \emph{International Conference on Network and
  Service Management (CNSM)}, 2024 (to appear).

\bibitem{qu2019delay}
K.~Qu, W.~Zhuang, Q.~Ye, X.~Shen, X.~Li, and J.~Rao, ``Delay-aware flow
  migration for embedded services in 5g core networks,'' in \emph{Proc. 2019
  IEEE International Conference on Communications (ICC)}, 2019, pp. 1--6.

\bibitem{luo2020effective}
J.~Luo, J.~Li, L.~Jiao, and J.~Cai, ``On the effective parallelization and
  near-optimal deployment of service function chains,'' \emph{IEEE Trans.
  Parallel Distrib. Syst.}, vol.~32, no.~5, pp. 1238--1255, 2020.

\bibitem{khosravian2020ietf}
P.~Khosravian, S.~Emadi, G.~Mirjalily, and B.~Zamani, ``Ietf-based finite
  automaton for service composition in service function chaining,''
  \emph{Wirel. Pers. Commun.}, vol. 114, no.~2, pp. 1235--1247, 2020.

\bibitem{cai2023privacy}
J.~Cai, Z.~Zhou, Z.~Huang, W.~Dai, and F.~R. Yu, ``Privacy-preserving
  deployment mechanism for service function chains across multiple domains,''
  \emph{IEEE Trans. Netw. Service Manag.}, vol.~21, no.~1, pp. 1241--1256,
  2024.

\bibitem{mehraghdam2016placement}
S.~Mehraghdam and H.~Karl, ``{Placement of services with flexible structures
  specified by a YANG data model},'' in \emph{Proc. IEEE NetSoft Conference and
  Workshops (NetSoft)}, 2016, pp. 184--192.

\bibitem{gilherrera2017scalable}
J.~Gil-Herrera and J.~F. Botero, ``A scalable metaheuristic for service
  function chain composition,'' in \emph{Proc. IEEE 9th Latin-American
  Conference on Communications (LATINCOM)}, 2017, pp. 1--6.

\bibitem{lawler1966branch}
E.~L. Lawler and D.~E. Wood, ``Branch-and-bound methods: A survey,''
  \emph{Oper. Res.}, vol.~14, no.~4, pp. 699--719, 1966.

\bibitem{morrison2016branch}
D.~R. Morrison, S.~H. Jacobson, J.~J. Sauppe, and E.~C. Sewell,
  ``Branch-and-bound algorithms: A survey of recent advances in searching,
  branching, and pruning,'' \emph{Discrete Optim.}, vol.~19, pp. 79--102, 2016.

\bibitem{wang2016branch}
Y.~Wang, Q.~Hu, and X.~Cao, ``A branch-and-price framework for optimal virtual
  network embedding,'' \emph{Comput. Netw.}, vol.~94, pp. 318--326, 2016.

\bibitem{taghavian2023cnsm}
M.~Taghavian, Y.~Hadjadj-Aoul, G.~Texier, N.~Huin, and P.~Bertin, ``A fair
  approach to the online placement of the network services over the edge,'' in
  \emph{Proc. 19th International Conference on Network and Service Management
  (CNSM)}, 2023, pp. 1--7.

\bibitem{taghavian2023approach}
------, ``An approach to network service placement reconciling optimality and
  scalability,'' \emph{IEEE Trans. Netw. Service Manag.}, vol.~20, no.~3, pp.
  2218--2229, 2023.

\bibitem{aklamanu2019utility}
F.~Aklamanu, S.~Randriamasy, and E.~Renault, ``Utility and a*-based algorithm
  for network slice placement and chaining,'' in \emph{Proc. 2019 IEEE Global
  Communications Conference (GLOBECOM)}, 2019, pp. 1--6.

\bibitem{luu2018aggregated}
Q.-T. Luu, M.~Kieffer, A.~Mouradian, and S.~Kerboeuf, ``Aggregated resource
  provisioning for network slices,'' in \emph{Proc. IEEE Global Communications
  Conference (GLOBECOM)}, 2018, pp. 1--6.

\bibitem{polese2023understanding}
M.~Polese, L.~Bonati, S.~D'Oro, S.~Basagni, and T.~Melodia, ``Understanding
  {O-RAN}: Architecture, interfaces, algorithms, security, and research
  challenges,'' \emph{IEEE Commun. Surv. Tutor.}, vol.~25, no.~2, pp.
  1376--1411, 2023.

\bibitem{savi2021impact}
M.~Savi, M.~Tornatore, and G.~Verticale, ``Impact of processing-resource
  sharing on the placement of chained virtual network functions,'' \emph{IEEE
  Trans. Cloud Comput.}, vol.~9, no.~4, pp. 1479--1492, 2021.

\bibitem{amaldi2016computational}
E.~Amaldi, S.~Coniglio, A.~M. Koster, and M.~Tieves, ``On the computational
  complexity of the virtual network embedding problem,'' \emph{Electron. Notes
  Discrete Math.}, vol.~52, pp. 213--220, 2016.

\bibitem{orlowski2010sndlib}
S.~Orlowski, R.~Wess{\"a}ly, M.~Pi{\'o}ro, and A.~Tomaszewski, ``{SNDlib}
  1.0-survivable network design library,'' \emph{Networks}, vol.~55, no.~3, pp.
  276--286, 2010.

\end{thebibliography}

\end{document}